\documentclass[%
 reprint,
 amsmath,amssymb,
 aps,
onecolumn,
]{revtex4-1}

\usepackage{graphicx}
\usepackage{dcolumn}
\usepackage{bm}
\usepackage{hyperref}

\usepackage{siunitx}
\graphicspath{{.}}
\DeclareGraphicsExtensions{.pdf,.png}

\newcommand{\beq}{\begin{IEEEeqnarray}{rCl}}
\newcommand{\beqr}{\begin{IEEEeqnarray}{rCl}}
\newcommand{\eeq}{\end{IEEEeqnarray}}
\newcommand{\eeqr}{\end{IEEEeqnarray}}
\newcommand{\mat}[1]{\mathbf{#1}}
\usepackage{IEEEtrantools}

\usepackage[nomargin,inline,final]{fixme}
\fxusetheme{color}

\begin{document}

\title{Neuromorphic Silicon Photonic Networks}

\author{Alexander~N.~Tait}
\email{atait@princeton.edu}
\author{Ellen~Zhou}
\author{Thomas~Ferreira~de~Lima}
\author{Allie~X.~Wu}
\author{Mitchell~A.~Nahmias}
\author{Bhavin~J.~Shastri}
\author{Paul~R.~Prucnal}

\affiliation{Princeton University, Princeton, NJ 08544, USA}%


\date{\today}

\begin{abstract}
Photonic systems for high-performance information processing have attracted renewed interest. Neuromorphic silicon photonics has the potential to integrate processing functions that vastly exceed the capabilities of electronics. We report first observations of a recurrent silicon photonic neural network, in which connections are configured by microring weight banks. A mathematical isomorphism between the silicon photonic circuit and a continuous neural network model is demonstrated through dynamical bifurcation analysis. Exploiting this isomorphism, a simulated 24-node silicon photonic neural network is programmed using ``neural compiler'' to solve a differential system emulation task. A 294-fold acceleration against a conventional benchmark is predicted. We also propose and derive power consumption analysis for modulator-class neurons that, as opposed to laser-class neurons, are compatible with silicon photonic platforms. At increased scale, Neuromorphic silicon photonics could access new regimes of ultrafast information processing for radio, control, and scientific computing.
\end{abstract}

\maketitle

    Light forms the global backbone of information transmission yet is rarely used for information transformation. Digital optical logic faces fundamental physical challenges~\cite{Keyes:1985}. Many analog approaches have been researched~\cite{Reimann:65,McCormick:93,Jutamulia1996}, but analog optical co-processors have faced major economic challenges. Optical systems have never achieved competitive manufacturability, nor have they satisfied a sufficiently general processing demand better than digital electronic contemporaries. Incipient changes in the supply and demand for photonics have the potential to spark a resurgence in optical information processing.

    A germinating silicon photonic integration industry promises to supply the manufacturing economomies normally reserved for microelectronics. While firmly rooted in demand for datacenter transceivers~\cite{Vlasov:12}, the industrialization of photonics would impact other application areas~\cite{Hochberg:13}. Industrial microfabrication ecosystems propel technology roadmapping~\cite{Thomson:16}, library standardization~\cite{Lim:14,Orcutt:12}, and broadened accessibility~\cite{Chrostowski:15}, all of which could open fundamentally new research directions into large-scale photonic systems. Large-scale beam steerers have been realized~\cite{Sun:14}, and on-chip communication networks have been envisioned~\cite{Beausoleil:2011,LeBeux:2011,Narayana:17}; however, opportunities for scalable silicon photonic information processing systems remain largely unexplored.

    Concurrently, photonic devices have found analog signal processing niches where electronics can no longer satisfy demands for bandwidth and reconfigurability. This situation is exemplified by radio frequency (RF) processing, in which front-ends have come to be limited by RF electronics, analog-to-digital converters (ADCs), and digital signal processors (DSP)~\cite{Capmany:13,Farsaei:16}. In response, RF photonics has offered respective solutions for tunable RF filters~\cite{Feng:10,Zhuang:15}, ADC itself~\cite{Valley:07}, and simple processing tasks that can be moved from DSP into the analog subsystem~\cite{Khan:2010,Chang:14implement,FerreiradeLima:16}. RF photonic circuits that can be transcribed from fiber to silicon are likely to reap the economic benefits of silicon photonic integration. In a distinct vein, an unprecedented possibility for large-scale photonic system integration could enable systems beyond what can be considered in fiber. If scalable information processing with analog photonics is to be considered, new standards relating physics to processing must be developed and verified.

    Standardized concepts that link physics to computational models are required to define essential quantitative engineering tools, namely metrics, algorithms, and benchmarks. For example, a conventional gate has simultaneous meaning as an abstract logical operation and as an arrangement of electronic semiconductors and thereby acts as a conduit between device engineering and computational performance. In another case, neuromorphic electronics adopt unconventional standards defining spiking neural networks as event-based packet networks~\cite{Merolla:2014,Akopyan:15,Indiveri:15}. These architectures' adherence to neural network models unlocks a wealth of metrics~\cite{Hasler2013}, algorithms~\cite{Wen:09,Lee:15}, tools~\cite{Eliasmith:04,Donnarumma:16}, and benchmarks~\cite{Diamond:16} developed specifically for neural networks. Likewise, scalable information processing with analog photonics would rely upon standards defining the relationship between photonic physics and a suitable processing model. Neural networks are among the most well-studied models for information processing with distributed analog elements. The fact that distributed interconnection and analog dynamics are performance strongsuits of photonic physics motivates the study of neuromorphic photonics.

\begin{figure}[ht]
        \centering
        \includegraphics[width=1.0\linewidth]{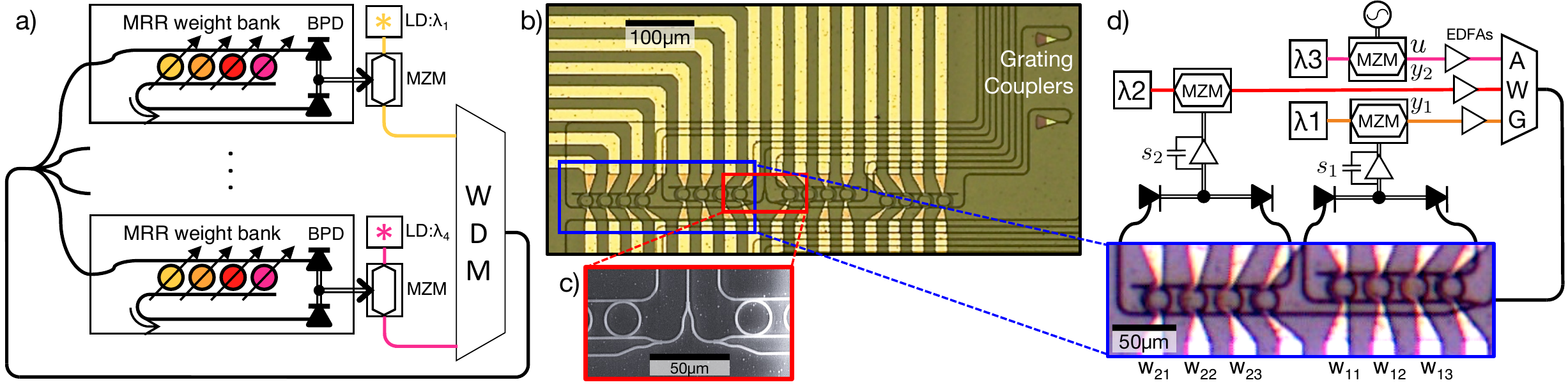}
        \caption{Broadcast-and-weight protocol and experiment. a) Concept of a broadcast-and-weight network with modulators used as neurons. MRR: microring resonator, BPD: balanced photodiode, LD: laser diode, MZM: Mach-Zehnder modulator, WDM: wavelength-division multiplexer. b) Micrograph of 4-node recurrent broadcast-and-weight network with 16 tunable microring (MRR) weights and fiber-to-chip grating couplers. c) Scanning electron micrograph of 1:4 splitter. d) Experimental setup with two off-chip MZM neurons and one external input. Signals are wavelength-multiplexed in an arrayed waveguide grating (AWG) and coupled into a 2x3 subnetwork with MRR weights, $w_{11}$, $w_{12}$, etc. Neuron state is represented by voltages $s_1$ and $s_2$ across low-pass filtered transimpedance amplifiers, which receive inputs from the balanced photodetectors of each MRR weight bank.}
        \label{fig:setup}
    \end{figure}

    ``Broadcast-and-weight''~\cite{Tait:14} was proposed as a standard protocol for implementing neuromorphic processors using integrated photonic devices. Broadcast-and-weight is potentially compatible with mainstream silicon photonic device platforms, unlike free-space optical neural networks~\cite{Jutamulia1996,Brunner:15}. It is capable of implementing generalized reconfigurable and recurrent neural network models. In the broadcast-and-weight protocol, shown in Fig.~\ref{fig:setup}(a), each neuron's output is assigned a unique wavelength carrier that is wavelength division multiplexed (WDM) and broadcast. Incoming WDM signals are weighted by reconfigurable, continuous-valued filters called photonic weight banks and then summed by total power detection. This electrical weighted sum then modulates the corresponding WDM carrier through a nonlinear or dynamical electro-optic process. Previous work on microring (MRR) weight banks have provided a standard for corresponding weighted addition operations and integrated photonic filters. In reference to the operation, MRR weight bank scalability~\cite{Tait:16scale} and accuracy~\cite{Tait:16multi} metrics can be defined, but MRR weight banks have not been demonstrated within a network.

    In this manuscript, we demonstrate a broadcast-and-weight system configured by microring weight banks that is isomorphic to a continuous-time recurrent neural network (CTRNN) model. As opposed to ``brain-inspired'' and ``neuro-mimetic,'' ``neuromorphic'' is an unambiguous mathematical concept meaning that a physical system's governing equations are isomorphic to those describing an abstract neural network model. Isomorphic dynamical systems share qualtitative changes in dynamics as a result of parameter variation. We adopt a strategy for proving neuromorphism experimentally by comparing dynamical transitions (a.k.a. bifurcations) induced in an experimental device against those predicted of a CTRNN model. In particular, we observe single-node bistability across a cusp bifurcation and two-node oscillation across a Hopf bifurcation. While oscillatory dynamics in optoelectronic devices have long been studied~\cite{Yamada:93,Romeira:14}, this work relies on configuring an analog photonic network that can be scaled to more nodes in principle. This implies that CTRNN metrics, simulators, algorithms, and benchmarks can be applied to larger neuromorphic silicon photonic systems. To illustrate the significance of this implication, we simulate a 24-modulator silicon photonic CTRNN solving a differential equation problem. The system is programmed by appropriating an existing ``neural compiler''~\cite{Eliasmith:04} and benchmarked against a conventional CPU solving the same problem, predicting an acceleration factor of 294$\times$.

\section*{Results}
    The CTRNN model is described by a set of ordinary differential equations coupled through a weight matrix.
    \beqr
    \frac{d\vec{s}(t)}{dt} &=& \mat{W} \vec{y}(t) - \frac{\vec{s}(t)}{\tau} + \vec{w}_{in} u(t) \label{eq:sysDefa}\\
    \vec{y}(t) &=& \sigma\left[\vec{s}(t)\right] \label{eq:sysDefb}
    \eeqr
    where $\vec{s}(t)$ are state variables with timeconstants $\tau$, $\mathbf{W}$ is the recurrent weight matrix, $\vec{y}(t)$ are neuron outputs, $\vec{w}_{in}$ are input weights, and $u(t)$ is an external input. $\sigma$ is a saturating transfer function associated with each neuron. Fig.~\ref{fig:setup}(b-c) shows the integrated, reconfigurable analog network and experimental setup. Signals $u$ and $\vec{y}$ are physically represented as the power envelope of different optical carrier wavelengths. The weight elements of $\mathbf{W}$ and $\vec{w}_{in}$ are implemented as transmission values through a network of reconfigurable MRR filters. The neuron transfer function, $\sigma$, is implemented by the sinusoidal electro-optic transfer function of a fiber Mach-Zehnder modulator (MZM). The neuron state, $s$, is the electrical voltage applied to the MZM, whose timeconstant, $\tau$, is determined by an electronic low-pass filter. We aim to establish a correspondence between experimental bifurcations induced by varying MRR weights and the modeled bifurcations~\cite{Beer:95} derived in Supplementary Section~1.

\begin{figure}[ht]
            \centering
            \includegraphics[width=.9\linewidth]{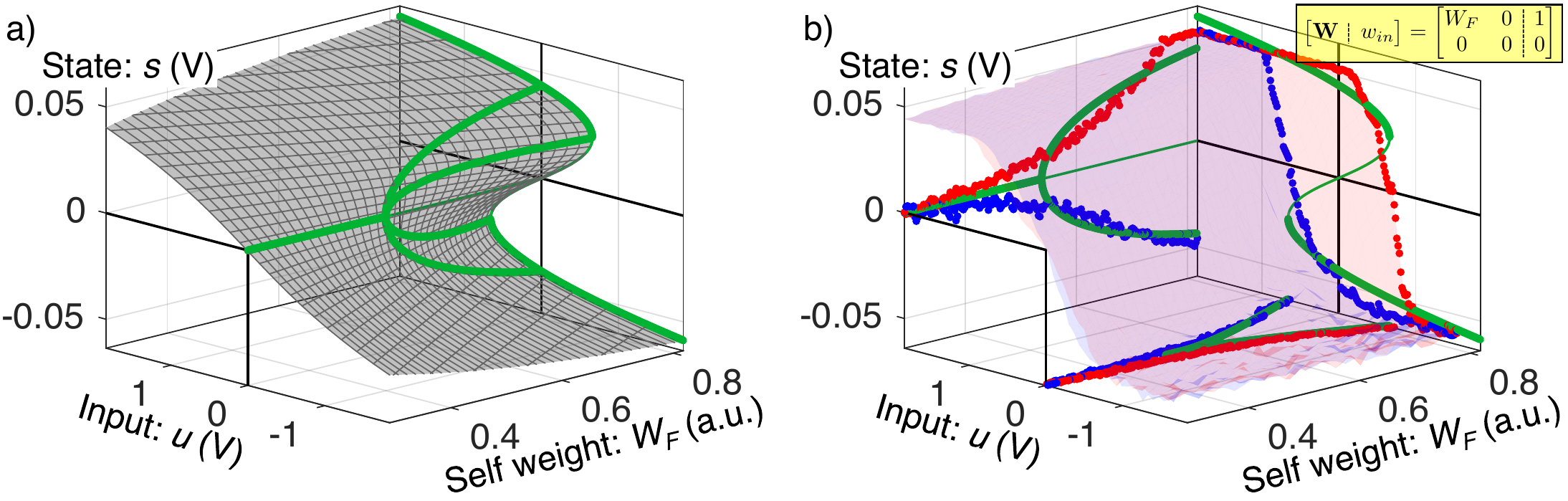}
            \caption{A cusp bifurcation in a single node with feedback weight, $W_F$, external input, $u$, and neurosynaptic state, $s$. a) Theoretical model surface (gray) and bifurcation curves (green) plotted in 3D. Parameters of the model are fit to data. b) Experimental data for increasing (blue surface) and decreasing (red surface) input. Theoretical bifurcation curves -- with parameters identical to those in (a) -- are projected onto 2D axes. The data surfaces are sliced at the planes: $u=0$, $s=0$, and $W_F=0.80$, and similarly projected onto the axes (red and blue points) to illustrate the reproduction of pitchfork, cusp, and saddle-node bifurcations, respectively}
            \label{fig:cusp}
        \end{figure}

    \paragraph{Cusp Bifurcation.}
        A cusp bifurcation characterizes the onset of bistability in a single node with self-feedback. To induce and observe a cusp, the feedback weight of node 1 is parameterized as $w_{11} = W_F$. The neurosynaptic signal, $s_1$, is recorded as the feedback weight is swept through 500 points from 0.05--0.85, and the input is swept in the rising (blue surface) and falling (red surface). The parameters $\tau$, $\alpha$, and $\kappa$ in equation~(S1.1) are fit to minimize root mean squared error between model and data surfaces. The best fit model has a cusp point at $W_B=0.54$. Fig.~\ref{fig:cusp}(a) shows a modeled cusp surface described by equation~(S1.1) and fit to data from Fig.~\ref{fig:cusp}(b). In Fig.~\ref{fig:cusp}(b), the data surfaces are interpolated at particular planes and projected onto 2D axes as red/blue points. The corresponding slices of the model surface are similarly projected as green lines. The $u=0$ slice projected on the $s$-$W_F$ plane yields a pitchfork curve described by equation~(S1.2). The $W_F=0.80$ slice projected on the $u$-$s$ yields the bistable curve described by equation~(S1.3). Finally, the $s=0$ slice projected on the $u$-$W_F$ plane yields the cusp curve described by equation~(S1.4).

        The experimental reproduction of pitchfork, bistable, and cusp bifurcations is demonstrative of an isomorphism between the single-node model and the device under test. An opening of an area between rising and falling data surfaces is characteristic of bistability. The transition boundary closely follows a cusp form. While the pitchfork and bistable slices reproduce the number and growth trends of fixed points, their fits have non-idealities. These non-idealities can be attributed to a hard saturation of the electrical transimpedance amplifier when the input voltage and feedback weight are high. Furthermore, the stability of cusp measurements serve as a control indicating the absence of time-delayed dynamics resulting from long fiber delays and causing spurious oscillations~\cite{Zhou:16oi}. Electrical low-pass filtering is used to eliminate these unmodeled dynamics in order to observe modeled bifurcations.

    \paragraph{Hopf Bifurcation.}
        Dynamical systems are capable of oscillating if there exists a closed orbit (a.k.a. limit cycle) in the state space, which must therefore exceed one dimension. The Hopf bifurcation occurs when a stable fixed-point becomes unstable while giving rise to a stable limit cycle. Hopf bifurcations are further characterized by oscillations that approach zero amplitude and nonzero frequency near the bifurcation point~\cite{Beer:95}. We induce a Hopf bifurcation by configuring the MRR weight matrix to have asymmetric, fixed off-diagonals and equal, parameterized diagonals. Reasoning for this choice is described in Methods.

        \begin{figure}[ht]
            \centering
            \includegraphics[width=.99\linewidth]{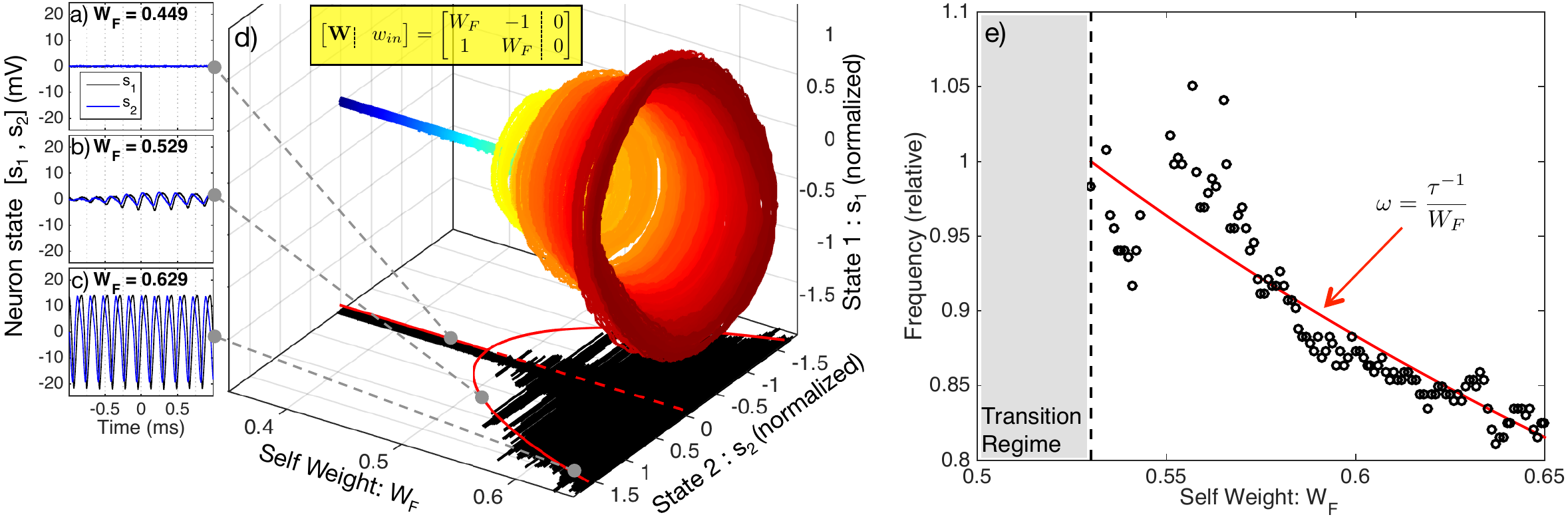}
            \caption{A Hopf bifurcation between stable and oscillating states. a-c) Time traces below, near, and above the bifurcation. d) Oscillation growth versus feedback weight strength. Color corresponds to feedback weight parameter, $W_F$, to improve visibility. Black shadow: average experimental amplitudes; solid red curve: corresponding fit model; dotted red line: unstable branch. e) Frequency of oscillation above the Hopf bifurcation. The observed data (black points) are compared to the expected trend of equation~(S1.10) (red curve). Frequencies are normalized to the threshold frequency of \SI{4.81}{k\Hz}.}
            \label{fig:hopf}
        \end{figure}

        Fig.~\ref{fig:hopf} compares the observed and predicted oscillation onset, amplitude, and frequency. Fig.~\ref{fig:hopf} (a-c) show the time traces for below, near, and above the oscillation threshold. Above threshold, oscillation occurs in the range of 1-5kHz, as limited by electronic low-pass filters and feedback delay. Fig.~\ref{fig:hopf}(d) shows the result of a fine sweep of self-feedback weights in the 2-node network, exhibiting the paraboloid shape of a Hopf bifurcation. The voltage of neuron 1 is plotted against that of neuron 2 with color corresponding to $W_F$ parameter. The peak oscillation amplitude for each weight is then projected onto the $W_F-y_2$ plane in black, and these amplitudes are fit using the model from equation~(S1.8) (red). Bifurcation occurs at $W_B=.48$ in the fit model. Fig.~\ref{fig:hopf}(e) plots the oscillation frequency above the Hopf point. Data are discarded for $W_B<W_F<0.53$ because the oscillations are erratic in the sensitive transition region. Frequency data are then fit with the model of equation~(S1.10). The frequency axis is scaled so that 1.0 corresponds to the model frequency at region boundary, which is \SI{4.81}{k\Hz}. The Hopf bifurcation only occurs in systems of more than one dimension, thus confirming the observation of a small integrated photonic neural network.

        Significantly above the bifurcation point, experimental oscillation amplitude and frequency closely match model predictions, but discrepancies are apparent in the transition regime. Limit cycles with amplitudes comparable to noise amplitude can be destabilized by their proximity to the unstable fixed point at zero. This effect could explain the middle inset of Fig.~\ref{fig:hopf}, in which a small oscillation grows and then shrinks. Part of this discrepancy can be explained by weight inaccuracy due to inter-bank thermal cross-talk. The two MRR weight banks were calibrated independently accounting only for intra-bank thermal cross-talk. As seen in Fig.~\ref{fig:setup}(c), the physical distance between $w_{12}$ (nominally --1) and $w_{22}$ (nominally $W_F$) is approximately \SI{100}{\um}. While inter-bank cross-talk is not a major effect, $w_{12}$ is very sensitive because weight --1 corresponds to on-resonance, and the dynamics are especially sensitive to the weight values near the bifurcation point.

\begin{figure}[ht]
            \centering
            \includegraphics[width=1.0\linewidth]{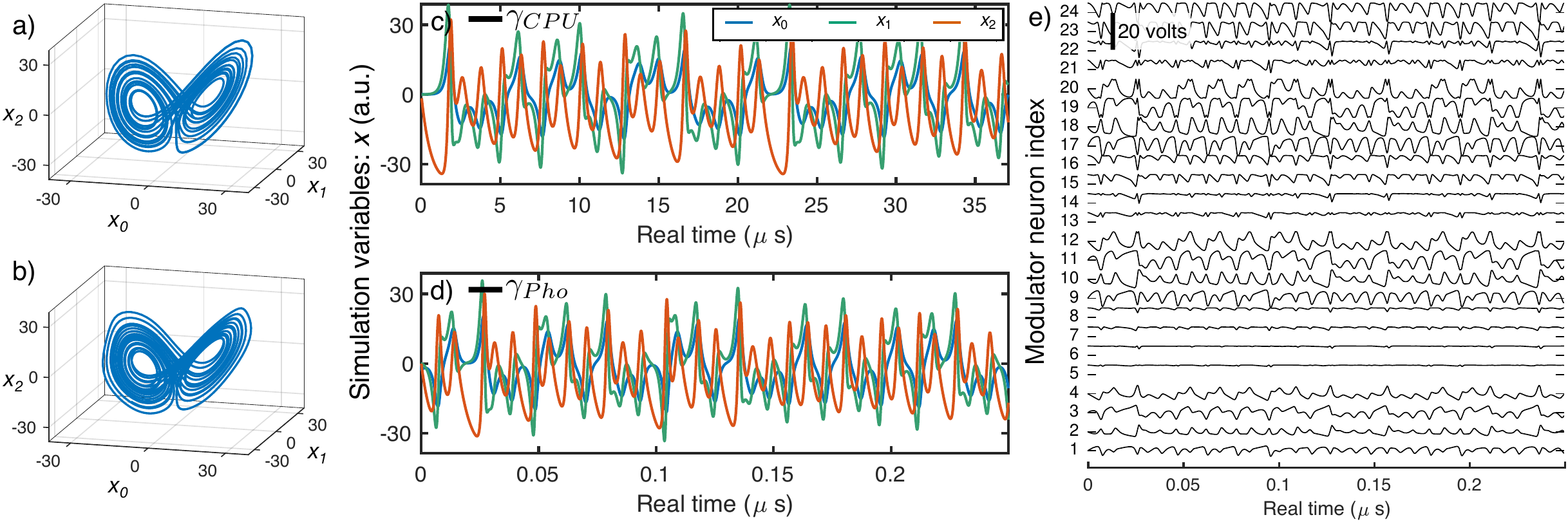}
            \caption{Photonic CTRNN benchmarking against a CPU. a-b) Phase diagrams of the Lorenz attractor simulated by a conventional CPU (a) and a photonic CTRNN (b). c-d) Time traces of simulation variables for a conventional CPU (c) and a photonic CTRNN (d). The horizontal axes are labeled in physical real time, and cover equal intervals of virtual simulation time, as benchmarked by $\gamma_{CPU}$ and $\gamma_{Pho}$. The ratio of real-time values of $\gamma$'s indicates a 294-fold acceleration. e) Time traces of modulator voltages $s_i$ (minor y-axis) for each modulator neuron $i$ (major y-axis) in the photonic CTRNN. The simulation variables, $\vec{x}$, in (d) are linear decodings of physical variables, $\vec{s}$, in (e).}
            \label{fig:lorenz}
        \end{figure}

    \paragraph{Emulation Benchmark.}
        A dynamical isomorphism between a silicon photonic system and the CTRNN model of equations~(\ref{eq:sysDefa}-\ref{eq:sysDefb}) implies that larger, faster neuromorphic silicon photonic systems could utilize algorithms and tools developed for generic CTRNNs. Here, we apply a ``neural compiler'' called the Neural Engineering Framework (NEF)~\cite{stewart2014large} to program a simulated photonic CTRNN to solve an ordinary differential equation (ODE). This simulation is benchmarked against a conventional central processing unit (CPU) solving the same task. The procedures for each approach are detailed in Methods. As opposed to implementation-specific metrics, benchmarks are task-oriented indicators suitable for comparing technologies using disparate computing standards. Benchmarking approaches are therefore needed to evaluate the potentials of any unconventional processor in application domains currently served by conventional processors. The chosen benchmark problem consists of solving a well-known ODE called the Lorenz attractor, described by a system of three coupled ODEs with no external inputs:
        \beqr
        \gamma \ \dot{x}_0 &=& \upsilon * (x_1 - x_0) \nonumber \\
        \gamma \ \dot{x}_1 &=& - x_0 x_2 - x_1 \label{eq:lorenz}\\
        \gamma \ \dot{x}_2 &=& x_0 x_1 - \beta (x_2 + \rho) - \rho \nonumber
        \eeqr
        where $\vec{x}$ are the simulation state variables, and $\gamma$ is a time scaling factor. When parameters, $\upsilon$, $\beta$, and $\rho$, are set to $(\upsilon, \beta, \rho) = (6.5, 8/3, 28)$, the solutions of the attractor are chaotic. The photonic CTRNN and CPU solutions are compared in $\vec{x}$ phase space in Fig.~\ref{fig:lorenz}(a-b) and the time-domain in Fig.~\ref{fig:lorenz}(c-d). Fig.~\ref{fig:lorenz}(e) plots the physical modulator voltages, $\vec{s}$, linear combinations of which represent simulation variables, $\vec{x}$, as discussed in Methods. Because the two simulators are implemented differently, they cannot be compared based on equivalent metrics; however, the time scaling factor, $\gamma$, links physical real-time to a virtual simulation time basis, in which a direct comparison can be made.

        Fig.~\ref{fig:lorenz}(c-d) plots photonic CTRNN and CPU solutions in the real-time bases, scaled to cover equal simulation intervals. The discrete-time simulation is linked to physical real-time by the step calculation time, $\Delta t =$ \SI{24.5}{\ns}, and its stability is limited by numerical divergence. We find that $\gamma_{CPU} \geq \Delta t \times 150$ is sufficient for $<$1\% divergence probability, resulting in $\gamma_{CPU} =$ \SI{3.68}{\us}. The CTRNN simulation is linked to physical real-time by its feedback delay, $t_{fb} =$ \SI{47.8}{\ps}, and its stability is limited by time-delayed dynamics. We find that $\gamma_{Pho} \geq t_{fb} \times 260$ is sufficient to avoid spurious dynamics, resulting in $\gamma_{Pho} =$ \SI{12.5}{\ns}. The acceleration factor, $\gamma_{CPU}/\gamma_{Pho}$, is thus predicted to be 294$\times$.

        Implementing this network on a silicon photonic chip would require 24 laser wavelengths, 24 modulators, and 24 MRR weight banks, for a total of 576 MRR weights. The power used by the CTRNN would be dominated by static MRR tuning and pump lasers,as discussed in the Methods section. Minimum pump power is determined by the requirement that recurrently connected neurons are able to exhibit positive eigenvalues. Considering 24 lasers with a realistic wall-plug efficiency of 5\%, minimum total system power is expected to be \SI{106}{\mW}. The area used by the photonic CTRNN is split evenly between MRR weight banks ($576 \times \SI{25}{\um} \times \SI{25}{\um} =$ \SI{0.36}{\mm^2}) and modulators~\cite{Khanna:15} ($24 \times \SI{500}{\um} \times \SI{25}{\um} =$ \SI{0.30}{\mm^2}). The fundamental limits of these performance metrics are compared with other neuromorphic approaches below.

        While the qualitative Lorenz behavior is reproduced by CTRNN and CPU implementations, the chaotic nature of the attractor presents a challenge for benchmarking emulation accuracy. Non-chaotic partial differential equations (PDEs) exist to serve as accuracy benchmarks~\cite{Roska:95,Ratier:12}; however, most non-trivial ODEs are chaotic. One exception is work on central pattern generators (CPGs) that are used to shape oscillations for driving mechanical locomotion~\cite{Vogelstein:08}. CPGs have been implemented with analog ASICs~\cite{Arena:05} and digital FPGAs~\cite{BarronZambrano:13}. While work on CPG hardware has fallen in sub-kHz timescales, similar tasks could be developed for GHz timescales with possible application to adaptive RF waveform generation. Further work could develop CPG-like tasks to benchmark the accuracy of photonic CTRNN ODE emulators.

        Accuracy can be assessed through metrics of weight accuracy. Previous work discussed the precision and accuracy to which MRR weight banks could be configured to a desired weight vector~\cite{Tait:16multi}. Even in the presence of fabrication variation and thermal cross-talk, the dynamic weight accuracy was demonstrated to be 4.1 + 1(sign) bits and anticipated to increase with improved design. Since the weight is analog, its dynamic accuracy (range / command error) is not an integer; however, this metric corresponds to bit resolution in digital architectures. In the majority of modern-day neuromorphic hardware, this resolution is selected to be 4 + 1(sign) bits~\cite{Akopyan:15,Friedmann:13,Benjamin:14} as a tradeoff between hardware resources and functionality. Significant study has been devoted to the effect of limited weight resolution on neural network function~\cite{Pfeil:12} and techniques for mitigating detrimental effects~\cite{Binas:16}.

    \paragraph{Neuromorphic photonic metrics}
        In addition to task-driven benchmark analyses, we can perform component-level metric comparisions with other neuromorphic systems in terms of area, signal bandwidth, and synaptic operation (SOP) energy.
        The 24-modulator photonic CTRNN used as an emulation benchmark was predicted to use a \SI{4.4}{\mW}/neuron using realistic pump lasers, and was limited to 1GHz bandwidth to spoil spurious oscillations. This results in a computational efficiency of 180 fJ/SOP.
        The area of an MRR weight is approximately $(l\times w)/N^2 = 25 \times 25 =$\SI{625}{\um^2}/synapse.

        The coherent approach described by Shen et al.~\cite{Shen:16arxiv} based on a matrix of Mach-Zehnder interferometers (MZIs) would exhibit similar fundamental energy and speed limitations, given similar assumptions about detection bandwidth and laser efficiency.
        The power requirements of this approach were limited by nonlinear threshold activation, rather than signal cascadability. Supposing a laser efficiency of 5\%, this was estimated to be around \SI{20}{\mW}/neuron. A 24-neuron system limited to 1GHz bandwidth would therefore achieve 830 fJ/SOP.
        While there is no one-to-one correspondence between MZIs and synapses, there is still a quadratic area scaling relationship: $(l\times w)/N^2 = 200 \times 100 =$20,000 \si{\um^2}/synapse, as limited by thermal phase shifter dimension.

        The superconducting optoelectronic approach described by Shainline et al.~\cite{Shainline:16} would be optimized for scalability and efficiency, instead of speed. This can be attributed to the extreme sensitivity of cryogenic photodetectors, but this difference also limits signal bandwidth to \SI{20}{\MHz}. For a 700-neuron interconnect, the wall-plug efficiency is estimated to be around \SI{20}fJ/SOP.
        Area was calculated to be $(l\times w)/N^2 = 1.4 \times 15 =$\SI{21}{\um^2}/synapse.

        A metric analysis can extend to include neuromorphic electronics, although metrics do not necessarily indicate signal processing merit. Electronic and photonic neuromorphics are designed to address complementary types of problems. Akopyan et al.~\cite{Akopyan:15} demonstrated a chip containing 256 million synapses dissipating \SI{65}{\mW}. The signal bandwidth, determined by the effective tick, or timestep, is \SI{1.0}{\kHz}, and the chip area is \SI{4.3}{\cm^2}. This results in an effective \SI{240}fJ/SOP and effective area of \SI{6.0}{\um^2}/synapse. We note that TrueNorth is event-based, meaning explicit computation does not occur for every effective SOP, but only when the input to a synapse is nonzero.

\section*{Discussion}
    We have demonstrated an isomorphism between a silicon photonic broadcast-and-weight system and a reconfigurable CTRNN model through observations of predicted bifurcations. In addition to proof-of-concept, this repeatable method could be used to characterize the standard performance of single neurons and pairs of neurons within larger systems. Employing neuromorphic properties, we then illustrated a task-oriented programming approach and benchmark analysis. Similar analyses could assess the potentials of analog photonic processors against state-of-the-art conventional processors in many application domains. Here, we discuss the implications of these results in the broader context of information processing with photonic devices.

        This work constitutes the first investigation of photonic neurons implemented by modulators, an important step towards silicon-compatible neuromorphic photonics. Interest in integrated lasers with neuron-like spiking behavior has flourished over the past several years~\cite{nahmias2013leaky,Prucnal:16advances}. Experimental work has so far focused on isolated neurons~\cite{Selmi:2014,Romeira:2016,Nahmias:16} and fixed, cascadable chains~\cite{VanVaerenbergh:12,Shastri:2015}. The shortage of research on networks of these lasers might be explained by the challenges of implementing low-loss, compact, and tunable filters in the active III/V platforms required for laser gain. In some cases where fan-in is sensitive to input optical phase, it is also unclear how networking would occur without global laser synchronization.
        In contrast to lasers, Mach-Zehnder, microring, and electroabsorption modulators are all silicon-compatible. Modulator-class neurons are therefore a final step towards making complete broadcast-and-weight systems entirely compatible with silicon foundry platforms. While laser-class neurons with spiking dynamics present richer processing opportunities, modulator-class neurons would still possess the formidable repertoire of CTRNN functions.

        In parallel with work on individual laser neurons, recent research has also investigated systems with isomorphisms to neural network models.
        A fully integrated superconducting optoelectronic network was recently proposed~\cite{Shainline:16} to offer unmatched energy efficiency. While based on an exotic superconducting platform, this approach accomplishes fan-in using incoherent optical power detection in a way compatible with the broadcast-and-weight protocol.
        A programmable nanophotonic processor was recently studied in the context of deep learning~\cite{Shen:16arxiv}. Coherent optical interconnects exhibit a sensitivity to optical phase that must be re-synchronized after each layer. In the demonstration, optical nonlinearity and phase regeneration were performed digitally. Analog solutions for counteracting signal-dependent phase shifts induced by nonlinear materials~\cite{Zhang:12} have not yet been proposed.
        Recurrent neural networks have been investigated in fiber~\cite{Hill:2002}. While the current work employs fiber neurons, it is the first demonstration of a recurrent weight network that is integrated. Optical neural networks in free-space have also been investigated in the past~\cite{Jutamulia1996} and recently~\cite{Brunner:15}. Free-space systems occupy an extra dimension but can not necessarily use it for increased scalability. The volume between focal planes is used for diffractive evolution of the optical field and unused for network configuration. Spatial light modulators that configure the network are generally planar. Shainline et al. noted that integrated neuromorphic photonic systems could potentially be stacked to take advantage of a third dimension~\cite{Shainline:16}.

        Reservoir computing techniques that take inspiration from certain brain properties (e.g. analog, distributed) have received substantial recent attention from the photonics community~\cite{Brunner:2013,Vandoorne:2014,Soriano:15,Duport:2016}. Reservoir techniques rely on supervised learning to discern a desired behavior from a large number of complex dynamics, instead of relying on establishing an isomorphism with a model.
        Neuromorphic and reservoir approaches differ fundamentally and possess complementary advantages. Both derive a broad repertoire of behaviors (often referred to as complexity) from a large number of physical degrees-of-freedom (e.g. intensities) coupled through interaction parameters (e.g. transmissions). Both offer means of selecting a specific, desired behavior from this repertoire using controllable parameters. In neuromorphic systems, network weights are \emph{both} the interaction and controllable parameters, whereas, in reservoir computers, these two groups of parameters are separate.
        This distinction has two major implications. Firstly, the interaction parameters of a reservoir do not need to be observable or even repeatable from system-to-system. Reservoirs can thus derive complexity from physical processes that are difficult to model or reproduce, such as coupled amplifiers~\cite{Vandoorne:08}, coupled nonlinear MRRs~\cite{Mesaritakis:13}, time-delayed dynamics in fibers~\cite{Soriano:15}, and fixed interferometric circuits~\cite{Vandoorne:2014}. Furthermore, they do not require significant hardware to control the state of the reservoir. Neuromorphic hardware has a burden to correspond physical parameters (e.g. drive voltages) to model parameters (e.g. weights), as was shown in this paper. Secondly, reservoir computers can only be made to elicit a desired behavior through instance-specific supervised training, whereas neuromorphic computers can be programmed \emph{a priori} using a known set of weights. Because neuromorphic behavior is determined only by controllable parameters, these parameters can be mapped directly between different system instances, different types of neuromorphic systems, and simulations. Neuromorphic hardware can leverage existing algorithms (e.g. NEF), map virtual training results to hardware, and particular behaviors are guaranteed to occur. Photonic RCs can of course be simulated; however, they have no corresponding guarantee that a particular hardware instance will reproduce a simulated behavior or that training will be able to converge to this behavior.

        At increased scale, neuromorphic silicon photonic systems could be applied to unaddressed computational areas in scientific computing and RF signal processing. A key benefit of neuromorphic engineering is that existing algorithms can be leveraged.
        A subset of CTRNNs, Hopfield networks~\cite{Hopfield:85}, have been used extensively in mathematical programming and optimization problems~\cite{Wen:09}. The ubiquity of PDE problems in scientific computing has motivated the development of analog electronic neural emulators~\cite{Roska:95}. Further work could explore the use of NEF to emulate discrete space points of PDEs.
        Neural algorithms for CTRNNs have been developed for real-time RF signal processing, including spectral mining~\cite{Tumuluru:10}, spread spectrum channel estimation~\cite{Mitra:94}, and arrayed antenna control~\cite{Du:02}. There is insistent demand to implement these tasks at wider bandwidths using less power than possible with RF electronics. Additionally, methodologies developed for audio applications, such as noise mitigation~\cite{Lee:15}, could conceivably be mapped to RF problems if implemented on ultrafast hardware. Unsupervised neural-inspired learning has been used with a single MRR weight bank for statistical analysis of multiple RF signals~\cite{Tait:17cleo}.

        We have demonstrated a reconfigurable analog neural network in a silicon photonic integrated circuit using modulators as neuron elements. Network-mediated cusp and Hopf bifurcations were observed as a proof-of-concept of an integrated broadcast-and-weight system~\cite{Tait:14}. Simulations of a 24 modulator neuron network performing an emulation task estimated a 294$\times$ speedup over a verified CPU benchmark. Neural network abstractions are powerful tools for bridging the gap between physical dynamics and useful application, and silicon photonic manufacturing introduces opportunities for large-scale photonic systems.

\section*{Methods}
    \paragraph{Experimental Setup.}
        Samples shown in Fig.~\ref{fig:setup}(b) were fabricated on silicon-on-insulator (SOI) wafers at the Washington Nanofabrication Facility through the SiEPIC Ebeam rapid prototyping group~\cite{Chrostowski:15}. Silicon thickness is \SI{220}{\nano\meter}, and buried oxide (BOX) thickness is \SI{3}{\micro\meter}. \SI{500}{\nano\meter} wide WGs were patterned by Ebeam lithography and fully etched through to the BOX~\cite{Bojko:11}. After a cladding oxide (\SI{3}{\um}) is deposited, Ti/W and Al layers are deposited. Ohmic heating in Ti/W filaments causes thermo-optic resonant wavelength shifts in the MRR weights. The sample is mounted on a temperature stabilized alignment stage and coupled to a 9-fiber array using focusing subwavelength grating couplers~\cite{Wang:14opex}. The reconfigurable analog network consists of 2 MRR weight banks each with four MRR weights with \SI{10}{\um} radii.

        Each MRR weight bank is calibrated using a multi-channel protocol described in past work~\cite{Tait:15cont,Tait:16multi}: an offline measurement procedure is performed to identify models of thermo-optic cross-talk and MRR filter edge transmission. During this calibration phase, electrical feedback connections are disabled and the set of wavelength channels carry a set of linearly seperable training signals. After calibration, the user can specify a desired weight matrix, and the control model calculates and applies the corresponding electrical currents.

        Weighted network outputs are detected off-chip, and the electrical weighted sums drive fiber Mach-Zehnder modulators (MZMs). Detected signals are low-pass filtered at 10kHz, represented by capacitor symbols in Fig.~\ref{fig:setup}(c). Low-pass filtering is used to spoil time-delayed dynamics that arise when feedback delay is much greater than the state time-constant~\cite{Romeira:14}. In this setup with on-chip network and off-chip modulator neurons, fiber delayed dynamics would interfere with CTRNN dynamical analysis~\cite{Zhou:16oi}. MZMs modulating distinct wavelengths $\lambda_1$=\SI{1549.97}{\nm} and $\lambda_2$=\SI{1551.68}{\nm} with neuron output signals $y_1(t)$ and $y_2(t)$, respectively. The MZM electro-optic transfer function serves as the nonlinear transfer function, $y=\sigma(s)$, associated with the continuous-time neuron. A third wavelength, $\lambda_3$=\SI{1553.46}{\nm}, carries an external input signal, $x(t)$, derived from a signal generator. Each laser diode source (ILX 7900B) outputs +13dBm of power. All optical signals ($u$, $y_1$, and $y_2$) are wavelength multiplexed in an arrayed waveguide grating (AWG) and then coupled back into the on-chip broadcast STAR consisting of splitting Y-junctions~\cite{Zhang:13} (Fig.~\ref{fig:setup}(c)).

        \paragraph{Photonic CTRNN Solver.}
            Recently developed compilers, such as Neural ENGineering Objects (Nengo)~\cite{Bekolay:2013}, employ the Neural Engineering Framework (NEF)~\cite{stewart2014large} to arrange networks of neurons to represent values and functions without relying on training. While originally developed to evaluate theories of cognition, the NEF has been appropriated to solve engineering problems~\cite{Friedl:16} and has been used to program electronic neuromorphic hardware~\cite{Mundy:15}. Background on the NEF compilation procedure is provided in Supplementary section~2. Simulation state variables, $\vec{x}$, are encoded as linear combinations of real population states, $\vec{s}$. Each neuron in a population has the same tuning curve shape, $\sigma$, but differ in gain $g$, input encoder vector $\vec{e}$, and offset $b$. The input-output relation of neuron $i$ -- equivalent to equation~\ref{eq:sysDefb}), is thus $s_i = \sigma(g_i \vec{e}_i \cdot \vec{x} + b_i)$. In this formulation, arbitrary nonlinear functions of the simulation variables, $\vec{f}\left(\vec{x}\right)$, are represented by linear combinations of the set of these tuning curves across the domain of values of $\vec{x}$ considered. Introducing recurrent connections in the population introduces the notion of state time-derivatives, as in equation~(\ref{eq:sysDefa}). By applying the decoder transform to both sides of equation~(\ref{eq:sysDefa}) and using the arbitrary function mapping technique to find $\mathbf{W}$, the neural population emulates an effective dynamical system of the form $\dot{\vec{x}} = \vec{f}\left(\vec{x}\right)$. Given equations~(\ref{eq:lorenz}) stated in this form, Nengo performs the steps necessary to represent the variables, functions, and complete ODE.

            Modifications were made to the standard Nengo procedure. Firstly, we specify the tuning curve shape as the sinusoidal electro-optic transfer characteristic of a MZM. Secondly, to reduce the number of MZMs required, we choose encoders to be the vertices of a unit-square $\{e\} = [1, \pm1, \pm1]$, while they are typically chosen randomly. Thirdly, the MZM sinusoidal transfer function provides a natural relation to the Fourier basis. Gains are chosen to correspond to the first three Fourier frequencies of the domain: $g\in s_\pi/2 \cdot \{1, 2, 3\}$, where $s_\pi$ is the MZM half-period. Offsets were chosen to be $b\in\{ 0,s_\pi/2\}$, corresponding to sine and cosine components of each gain frequency. The total number of modulator neurons is therefore $\# e \cdot \# g \cdot \# b = 4\cdot3\cdot2 = 24$. Fig.~\ref{fig:lorenz}(e) shows the MZM states, $s(t)$, of which simulation variables, $x(t)$, are linear combinations. From this plot, it appears that some neurons are barely used. Thus, further optimizations of number of neurons could be made by pruning those neurons after compilation of the weight matrix.

            The operational speed of this network would be limited by time-of-flight feedback delay. In Fig.~\ref{fig:setup}(a), the longest feedback path is via the drop port of the last (pink) MRR weight of the first (yellow) neuron's bank. The path includes the perimeter of the square MRR weight matrix, plus a drop waveguide folded back along the bank. Supposing a minimum MRR pitch of \SI{25}{\um} and MZM length of \SI{500}{\um}, the feedback delay would then be $\left(6 \times 25 \times 25 + 500\right) \cdot n/c =$ \SI{48}{\ps}. We model this delayed feedback in the Nengo simulation and then adjust feedback strength to find the minimum stable simulation timescale. For $\gamma_{Pho} / t_{fb} <$ 65, spurious time-delayed dynamics dominate. For $\gamma_{Pho} / t_{fb} <$ 104, the butterfly phase diagram in Fig.~\ref{fig:lorenz}(b) is not reproduced accurately. $\gamma_{Pho} / t_{fb} \geq$ 260 is chosen for robust reproduction of the expected dynamics.

        \paragraph{Conventional CPU Solver.}
            Conventional digital processors must use a discrete-time approximation to simulate continuous ODEs, the simplest of which is Euler continuation:
            \beq
            \vec{x}[(n+1) \Delta t] = \vec{x}[n\Delta t] + \Delta t \vec{f}(\vec{x}[n\Delta t])
            \label{eq:euler}
            \eeq
            where $\Delta t$ is the time step interval. To estimate the real-time value of $\Delta t$, we develop and validate a simple CPU model. For each time step, the CPU must compute $\vec{f}(\vec{x}[n\Delta t])$ as defined in equations~(\ref{eq:lorenz}), resulting in 9 floating-point operations (FLOPs), and 12 cache reads of the operands. The Euler update in equation~(\ref{eq:euler}) constitutes one multiply, one addition, and one read/write for each state variable, resulting in 6 FLOPs and 6 cache accesses. Supposing a FLOP latency of 1 clock cycle, Level 1 (L1) cache latency of 4 cycles, and 2.6GHz clock, this model predicts a time step of $\Delta t = $\SI{33}{\ns}. This model is empirically validated using an Intel Core i5-4288U. The machine-optimized program randomly initializes and loops through $10^6$ Euler steps of the Lorenz system, over 100 trials. CPU time was measured to be $\Delta t = 24.5 \pm 1.5$ns. The minimum stable simulation timescale is limited by divergent errors stemming from time discretization. We performed a series of 100 trials over 100 values of $\Delta t / \gamma_{CPU}$, finding that $<$1\% probability of divergence occurred for $\gamma_{CPU} / \Delta t \geq$ 150.

    \paragraph{Minimum Power Calculations}
        Static thermal power must be applied to each weight in order to track MRRs to the on-resonance condition. Supposing a bank length set by an MRR pitch of \SI{25}{\um} and count of 24, the MRR network would occupy a square with \SI{600}{\um} sides. Within this length, resonances can be fabricated with repeatability within $\pm$1.3nm~\cite{Chrostowski:14}. Supposing a tuning efficiency of \SI{0.25}{\nm}/mW~\cite{Jayatilleka:15opex}, it would take an average of \SI{5.2}{\mW}/weight to track resonance, for a static power dissipation of \SI{3.0}{\W}. On the other hand, if depletion-based tuning can be used, there would be negligible static power dissipation in the weights.

        The laser bias power must be set such that a modulator neuron can drive downstream neurons with sufficient strength. A neuron fed back to itself should be able to elicit an equivalent or greater small-signal response after one round-trip. This condition is referred to as signal cascadability and can be stated as $g\geq1$, where $g$ is round-trip, small-signal gain. If the cascadability condition is not met, all signals will eventually attenuate out with time. In a recurrent network, the real part of system eigenvalues would not be able to exceed zero. Round-trip gain is expressed as
        \beq
        g = \frac{dP_{out}}{dP_{in}}
        \eeq
        For a modulator-based broadcast-and-weight system, this breaks down into receiver and modulator components. Assuming a voltage-mode modulator, such as reverse-biased MRR depletion modulator,
        \beqr
        \left.\frac{dP_{out}}{dV_{mod}}\right|_{max} &=& \frac{\pi}{2 V_{\pi}} P_{pump}\\
        \frac{dV_{mod}}{dP_{in}} &=& R_{PD} R_r
        \eeqr
        Where $V_\pi$ is modulator $\pi$-voltage, $P_{pump}$ is modulator pump power, and $R_{PD}$ is detector responsivity. Because input power generates a photocurrent, yet a depletion modulator is voltage-driven. The receiver's impedance, $R_r$, determines the conversion and can be set externally. As $R_r$ increases, round-trip gain also increases, but bandwidth decreases according to $f = (2 \pi R_r C_{mod})^{-1}$, where $C_{mod}$ is PN junction capacitance of the modulator. By setting the cascadability condition: $g=1$ and combining the above equations, we find that
        \beqr
        P_{pump}(R_r) &=& \frac{2 V_\pi}{\pi R_{PD} R_r} \\
        P_{pump}(f) &=& 4 (V_\pi C_{mod}) R_{PD}^{-1} f
        \eeqr
        The values of $V_\pi$, $C_{mod}$, and $R_{PD}$ on a typical silicon photonic foundry platform have been published~\cite{Khanna:15}. For an MRR depletion modulator, $V_\pi = 1.5$ V, $C_{mod} = 35$ fF. For a PD on the same platform, $R_{PD} = 0.97$ A/W. This means that the minimum pump power for a given signal bandwidth is $2.2 \times 10^{-13}$ W/Hz.

        In this paper, we study a 24-node CTRNN whose signal bandwidth is restricted to 1GHz to avoid time-delay dynamics. This means that, for the cascadability condition to be met, modulator pumping must be at least \SI{0.22}{\mW}/neuron of optical power. Adding up 24 lasers and accounting for laser inefficiency, wall-plug system power would be \SI{106}{\mW}.

\begin{acknowledgments}
    This work is supported by National Science Foundation (NSF) Enhancing Access to the Radio Spectrum (EARS) program (Award 1642991). Fabrication support was provided via the Natural Sciences and Engineering Research Council of Canada (NSERC) Silicon Electronic-Photonic Integrated Circuits (SiEPIC) Program. Devices were fabricated by Richard Bojko at the University of Washington Washington Nanofabrication Facility, part of the NSF National Nanotechnology Infrastructure Network (NNIN).
\end{acknowledgments}




\begin{thebibliography}{81}%
\makeatletter
\providecommand \@ifxundefined [1]{%
 \@ifx{#1\undefined}
}%
\providecommand \@ifnum [1]{%
 \ifnum #1\expandafter \@firstoftwo
 \else \expandafter \@secondoftwo
 \fi
}%
\providecommand \@ifx [1]{%
 \ifx #1\expandafter \@firstoftwo
 \else \expandafter \@secondoftwo
 \fi
}%
\providecommand \natexlab [1]{#1}%
\providecommand \enquote  [1]{``#1''}%
\providecommand \bibnamefont  [1]{#1}%
\providecommand \bibfnamefont [1]{#1}%
\providecommand \citenamefont [1]{#1}%
\providecommand \href@noop [0]{\@secondoftwo}%
\providecommand \href [0]{\begingroup \@sanitize@url \@href}%
\providecommand \@href[1]{\@@startlink{#1}\@@href}%
\providecommand \@@href[1]{\endgroup#1\@@endlink}%
\providecommand \@sanitize@url [0]{\catcode `\\12\catcode `\$12\catcode
  `\&12\catcode `\#12\catcode `\^12\catcode `\_12\catcode `\%12\relax}%
\providecommand \@@startlink[1]{}%
\providecommand \@@endlink[0]{}%
\providecommand \url  [0]{\begingroup\@sanitize@url \@url }%
\providecommand \@url [1]{\endgroup\@href {#1}{\urlprefix }}%
\providecommand \urlprefix  [0]{URL }%
\providecommand \Eprint [0]{\href }%
\providecommand \doibase [0]{http://dx.doi.org/}%
\providecommand \selectlanguage [0]{\@gobble}%
\providecommand \bibinfo  [0]{\@secondoftwo}%
\providecommand \bibfield  [0]{\@secondoftwo}%
\providecommand \translation [1]{[#1]}%
\providecommand \BibitemOpen [0]{}%
\providecommand \bibitemStop [0]{}%
\providecommand \bibitemNoStop [0]{.\EOS\space}%
\providecommand \EOS [0]{\spacefactor3000\relax}%
\providecommand \BibitemShut  [1]{\csname bibitem#1\endcsname}%
\let\auto@bib@innerbib\@empty
\bibitem [{\citenamefont {Keyes}(1985)}]{Keyes:1985}%
  \BibitemOpen
  \bibfield  {author} {\bibinfo {author} {\bibfnamefont {R.~W.}\ \bibnamefont
  {Keyes}},\ }\href {\doibase 10.1080/713821757} {\bibfield  {journal}
  {\bibinfo  {journal} {Optica Acta: International Journal of Optics}\ }\textbf
  {\bibinfo {volume} {32}},\ \bibinfo {pages} {525} (\bibinfo {year} {1985})},\
  \Eprint
  {http://arxiv.org/abs/http://www.tandfonline.com/doi/pdf/10.1080/713821757}
  {http://www.tandfonline.com/doi/pdf/10.1080/713821757} \BibitemShut {NoStop}%
\bibitem [{\citenamefont {Reimann}\ and\ \citenamefont
  {Kosonocky}(1965)}]{Reimann:65}%
  \BibitemOpen
  \bibfield  {author} {\bibinfo {author} {\bibfnamefont {O.~A.}\ \bibnamefont
  {Reimann}}\ and\ \bibinfo {author} {\bibfnamefont {W.~F.}\ \bibnamefont
  {Kosonocky}},\ }\href {\doibase 10.1109/MSPEC.1965.5531775} {\bibfield
  {journal} {\bibinfo  {journal} {IEEE Spectrum}\ }\textbf {\bibinfo {volume}
  {2}},\ \bibinfo {pages} {181} (\bibinfo {year} {1965})}\BibitemShut {NoStop}%
\bibitem [{\citenamefont {McCormick}\ \emph {et~al.}(1993)\citenamefont
  {McCormick}, \citenamefont {Cloonan}, \citenamefont {Tooley}, \citenamefont
  {Lentine}, \citenamefont {Sasian}, \citenamefont {Brubaker}, \citenamefont
  {Morrison}, \citenamefont {Walker}, \citenamefont {Crisci}, \citenamefont
  {Novotny}, \citenamefont {Hinterlong}, \citenamefont {Hinton},\ and\
  \citenamefont {Kerbis}}]{McCormick:93}%
  \BibitemOpen
  \bibfield  {author} {\bibinfo {author} {\bibfnamefont {F.~B.}\ \bibnamefont
  {McCormick}}, \bibinfo {author} {\bibfnamefont {T.~J.}\ \bibnamefont
  {Cloonan}}, \bibinfo {author} {\bibfnamefont {F.~A.~P.}\ \bibnamefont
  {Tooley}}, \bibinfo {author} {\bibfnamefont {A.~L.}\ \bibnamefont {Lentine}},
  \bibinfo {author} {\bibfnamefont {J.~M.}\ \bibnamefont {Sasian}}, \bibinfo
  {author} {\bibfnamefont {J.~L.}\ \bibnamefont {Brubaker}}, \bibinfo {author}
  {\bibfnamefont {R.~L.}\ \bibnamefont {Morrison}}, \bibinfo {author}
  {\bibfnamefont {S.~L.}\ \bibnamefont {Walker}}, \bibinfo {author}
  {\bibfnamefont {R.~J.}\ \bibnamefont {Crisci}}, \bibinfo {author}
  {\bibfnamefont {R.~A.}\ \bibnamefont {Novotny}}, \bibinfo {author}
  {\bibfnamefont {S.~J.}\ \bibnamefont {Hinterlong}}, \bibinfo {author}
  {\bibfnamefont {H.~S.}\ \bibnamefont {Hinton}}, \ and\ \bibinfo {author}
  {\bibfnamefont {E.}~\bibnamefont {Kerbis}},\ }\href {\doibase
  10.1364/AO.32.005153} {\bibfield  {journal} {\bibinfo  {journal} {Appl.
  Opt.}\ }\textbf {\bibinfo {volume} {32}},\ \bibinfo {pages} {5153} (\bibinfo
  {year} {1993})}\BibitemShut {NoStop}%
\bibitem [{\citenamefont {Jutamulia}\ and\ \citenamefont
  {Yu}(1996)}]{Jutamulia1996}%
  \BibitemOpen
  \bibfield  {author} {\bibinfo {author} {\bibfnamefont {S.}~\bibnamefont
  {Jutamulia}}\ and\ \bibinfo {author} {\bibfnamefont {F.}~\bibnamefont {Yu}},\
  }\href {\doibase http://dx.doi.org/10.1016/0030-3992(95)00070-4} {\bibfield
  {journal} {\bibinfo  {journal} {Optics \& Laser Technology}\ }\textbf
  {\bibinfo {volume} {28}},\ \bibinfo {pages} {59 } (\bibinfo {year}
  {1996})}\BibitemShut {NoStop}%
\bibitem [{\citenamefont {Vlasov}(2012)}]{Vlasov:12}%
  \BibitemOpen
  \bibfield  {author} {\bibinfo {author} {\bibfnamefont {Y.}~\bibnamefont
  {Vlasov}},\ }\href {\doibase 10.1109/MCOM.2012.6146487} {\bibfield  {journal}
  {\bibinfo  {journal} {IEEE Commun. Mag.}\ }\textbf {\bibinfo {volume} {50}},\
  \bibinfo {pages} {s67} (\bibinfo {year} {2012})}\BibitemShut {NoStop}%
\bibitem [{\citenamefont {Hochberg}\ \emph {et~al.}(2013)\citenamefont
  {Hochberg}, \citenamefont {Harris}, \citenamefont {Ding}, \citenamefont
  {Zhang}, \citenamefont {Novack}, \citenamefont {Xuan},\ and\ \citenamefont
  {Baehr-Jones}}]{Hochberg:13}%
  \BibitemOpen
  \bibfield  {author} {\bibinfo {author} {\bibfnamefont {M.}~\bibnamefont
  {Hochberg}}, \bibinfo {author} {\bibfnamefont {N.~C.}\ \bibnamefont
  {Harris}}, \bibinfo {author} {\bibfnamefont {R.}~\bibnamefont {Ding}},
  \bibinfo {author} {\bibfnamefont {Y.}~\bibnamefont {Zhang}}, \bibinfo
  {author} {\bibfnamefont {A.}~\bibnamefont {Novack}}, \bibinfo {author}
  {\bibfnamefont {Z.}~\bibnamefont {Xuan}}, \ and\ \bibinfo {author}
  {\bibfnamefont {T.}~\bibnamefont {Baehr-Jones}},\ }\href {\doibase
  10.1109/MSSC.2012.2232791} {\bibfield  {journal} {\bibinfo  {journal} {IEEE
  Solid-State Circuits Magazine}\ }\textbf {\bibinfo {volume} {5}},\ \bibinfo
  {pages} {48} (\bibinfo {year} {2013})}\BibitemShut {NoStop}%
\bibitem [{\citenamefont {Thomson}\ \emph {et~al.}(2016)\citenamefont
  {Thomson}, \citenamefont {Zilkie}, \citenamefont {Bowers}, \citenamefont
  {Komljenovic}, \citenamefont {Reed}, \citenamefont {Vivien}, \citenamefont
  {Marris-Morini}, \citenamefont {Cassan}, \citenamefont {Virot}, \citenamefont
  {F{\'e}d{\'e}li}, \citenamefont {Hartmann}, \citenamefont {Schmid},
  \citenamefont {Xu}, \citenamefont {Boeuf}, \citenamefont {O'Brien},
  \citenamefont {Mashanovich},\ and\ \citenamefont {Nedeljkovic}}]{Thomson:16}%
  \BibitemOpen
  \bibfield  {author} {\bibinfo {author} {\bibfnamefont {D.}~\bibnamefont
  {Thomson}}, \bibinfo {author} {\bibfnamefont {A.}~\bibnamefont {Zilkie}},
  \bibinfo {author} {\bibfnamefont {J.~E.}\ \bibnamefont {Bowers}}, \bibinfo
  {author} {\bibfnamefont {T.}~\bibnamefont {Komljenovic}}, \bibinfo {author}
  {\bibfnamefont {G.~T.}\ \bibnamefont {Reed}}, \bibinfo {author}
  {\bibfnamefont {L.}~\bibnamefont {Vivien}}, \bibinfo {author} {\bibfnamefont
  {D.}~\bibnamefont {Marris-Morini}}, \bibinfo {author} {\bibfnamefont
  {E.}~\bibnamefont {Cassan}}, \bibinfo {author} {\bibfnamefont
  {L.}~\bibnamefont {Virot}}, \bibinfo {author} {\bibfnamefont {J.-M.}\
  \bibnamefont {F{\'e}d{\'e}li}}, \bibinfo {author} {\bibfnamefont {J.-M.}\
  \bibnamefont {Hartmann}}, \bibinfo {author} {\bibfnamefont {J.~H.}\
  \bibnamefont {Schmid}}, \bibinfo {author} {\bibfnamefont {D.-X.}\
  \bibnamefont {Xu}}, \bibinfo {author} {\bibfnamefont {F.}~\bibnamefont
  {Boeuf}}, \bibinfo {author} {\bibfnamefont {P.}~\bibnamefont {O'Brien}},
  \bibinfo {author} {\bibfnamefont {G.~Z.}\ \bibnamefont {Mashanovich}}, \ and\
  \bibinfo {author} {\bibfnamefont {M.}~\bibnamefont {Nedeljkovic}},\ }\href
  {\doibase 10.1088/2040-8978/18} {\bibfield  {journal} {\bibinfo  {journal}
  {Journal of Optics}\ }\textbf {\bibinfo {volume} {18}},\ \bibinfo {pages}
  {073003} (\bibinfo {year} {2016})}\BibitemShut {NoStop}%
\bibitem [{\citenamefont {Lim}\ \emph {et~al.}(2014)\citenamefont {Lim},
  \citenamefont {Song}, \citenamefont {Fang}, \citenamefont {Li}, \citenamefont
  {Tu}, \citenamefont {Duan}, \citenamefont {Chen}, \citenamefont {Tern},\ and\
  \citenamefont {Liow}}]{Lim:14}%
  \BibitemOpen
  \bibfield  {author} {\bibinfo {author} {\bibfnamefont {A.-J.}\ \bibnamefont
  {Lim}}, \bibinfo {author} {\bibfnamefont {J.}~\bibnamefont {Song}}, \bibinfo
  {author} {\bibfnamefont {Q.}~\bibnamefont {Fang}}, \bibinfo {author}
  {\bibfnamefont {C.}~\bibnamefont {Li}}, \bibinfo {author} {\bibfnamefont
  {X.}~\bibnamefont {Tu}}, \bibinfo {author} {\bibfnamefont {N.}~\bibnamefont
  {Duan}}, \bibinfo {author} {\bibfnamefont {K.~K.}\ \bibnamefont {Chen}},
  \bibinfo {author} {\bibfnamefont {R.-C.}\ \bibnamefont {Tern}}, \ and\
  \bibinfo {author} {\bibfnamefont {T.-Y.}\ \bibnamefont {Liow}},\ }\href
  {\doibase 10.1109/JSTQE.2013.2293274} {\bibfield  {journal} {\bibinfo
  {journal} {IEEE J. Sel. Top. Quantum Electron.}\ }\textbf {\bibinfo {volume}
  {20}},\ \bibinfo {pages} {405} (\bibinfo {year} {2014})}\BibitemShut
  {NoStop}%
\bibitem [{\citenamefont {Orcutt}\ \emph {et~al.}(2012)\citenamefont {Orcutt},
  \citenamefont {Moss}, \citenamefont {Sun}, \citenamefont {Leu}, \citenamefont
  {Georgas}, \citenamefont {Shainline}, \citenamefont {Zgraggen}, \citenamefont
  {Li}, \citenamefont {Sun}, \citenamefont {Weaver}, \citenamefont
  {Uro\v{s}evi\'{c}}, \citenamefont {Popovi\'{c}}, \citenamefont {Ram},\ and\
  \citenamefont {Stojanovi\'{c}}}]{Orcutt:12}%
  \BibitemOpen
  \bibfield  {author} {\bibinfo {author} {\bibfnamefont {J.~S.}\ \bibnamefont
  {Orcutt}}, \bibinfo {author} {\bibfnamefont {B.}~\bibnamefont {Moss}},
  \bibinfo {author} {\bibfnamefont {C.}~\bibnamefont {Sun}}, \bibinfo {author}
  {\bibfnamefont {J.}~\bibnamefont {Leu}}, \bibinfo {author} {\bibfnamefont
  {M.}~\bibnamefont {Georgas}}, \bibinfo {author} {\bibfnamefont
  {J.}~\bibnamefont {Shainline}}, \bibinfo {author} {\bibfnamefont
  {E.}~\bibnamefont {Zgraggen}}, \bibinfo {author} {\bibfnamefont
  {H.}~\bibnamefont {Li}}, \bibinfo {author} {\bibfnamefont {J.}~\bibnamefont
  {Sun}}, \bibinfo {author} {\bibfnamefont {M.}~\bibnamefont {Weaver}},
  \bibinfo {author} {\bibfnamefont {S.}~\bibnamefont {Uro\v{s}evi\'{c}}},
  \bibinfo {author} {\bibfnamefont {M.}~\bibnamefont {Popovi\'{c}}}, \bibinfo
  {author} {\bibfnamefont {R.~J.}\ \bibnamefont {Ram}}, \ and\ \bibinfo
  {author} {\bibfnamefont {V.}~\bibnamefont {Stojanovi\'{c}}},\ }\href
  {\doibase 10.1364/OE.20.012222} {\bibfield  {journal} {\bibinfo  {journal}
  {Opt. Express}\ }\textbf {\bibinfo {volume} {20}},\ \bibinfo {pages} {12222}
  (\bibinfo {year} {2012})}\BibitemShut {NoStop}%
\bibitem [{\citenamefont {Chrostowski}\ and\ \citenamefont
  {Hochberg}(2015)}]{Chrostowski:15}%
  \BibitemOpen
  \bibfield  {author} {\bibinfo {author} {\bibfnamefont {L.}~\bibnamefont
  {Chrostowski}}\ and\ \bibinfo {author} {\bibfnamefont {M.}~\bibnamefont
  {Hochberg}},\ }\href@noop {} {\emph {\bibinfo {title} {Silicon Photonics
  Design: From Devices to Systems}}}\ (\bibinfo  {publisher} {Cambridge
  University Press},\ \bibinfo {year} {2015})\BibitemShut {NoStop}%
\bibitem [{\citenamefont {Sun}\ \emph {et~al.}(2014)\citenamefont {Sun},
  \citenamefont {Timurdogan}, \citenamefont {Yaacobi}, \citenamefont {Su},
  \citenamefont {Hosseini}, \citenamefont {Cole},\ and\ \citenamefont
  {Watts}}]{Sun:14}%
  \BibitemOpen
  \bibfield  {author} {\bibinfo {author} {\bibfnamefont {J.}~\bibnamefont
  {Sun}}, \bibinfo {author} {\bibfnamefont {E.}~\bibnamefont {Timurdogan}},
  \bibinfo {author} {\bibfnamefont {A.}~\bibnamefont {Yaacobi}}, \bibinfo
  {author} {\bibfnamefont {Z.}~\bibnamefont {Su}}, \bibinfo {author}
  {\bibfnamefont {E.}~\bibnamefont {Hosseini}}, \bibinfo {author}
  {\bibfnamefont {D.}~\bibnamefont {Cole}}, \ and\ \bibinfo {author}
  {\bibfnamefont {M.}~\bibnamefont {Watts}},\ }\href {\doibase
  10.1109/JSTQE.2013.2293316} {\bibfield  {journal} {\bibinfo  {journal}
  {Selected Topics in Quantum Electronics, IEEE Journal of}\ }\textbf {\bibinfo
  {volume} {20}},\ \bibinfo {pages} {264} (\bibinfo {year} {2014})}\BibitemShut
  {NoStop}%
\bibitem [{\citenamefont {Beausoleil}(2011)}]{Beausoleil:2011}%
  \BibitemOpen
  \bibfield  {author} {\bibinfo {author} {\bibfnamefont {R.~G.}\ \bibnamefont
  {Beausoleil}},\ }\href {\doibase 10.1145/1970406.1970408} {\bibfield
  {journal} {\bibinfo  {journal} {J. Emerg. Technol. Comput. Syst.}\ }\textbf
  {\bibinfo {volume} {7}},\ \bibinfo {pages} {6:1} (\bibinfo {year}
  {2011})}\BibitemShut {NoStop}%
\bibitem [{\citenamefont {Le~Beux}\ \emph {et~al.}(2011)\citenamefont
  {Le~Beux}, \citenamefont {Trajkovic}, \citenamefont {O'Connor}, \citenamefont
  {Nicolescu}, \citenamefont {Bois},\ and\ \citenamefont
  {Paulin}}]{LeBeux:2011}%
  \BibitemOpen
  \bibfield  {author} {\bibinfo {author} {\bibfnamefont {S.}~\bibnamefont
  {Le~Beux}}, \bibinfo {author} {\bibfnamefont {J.}~\bibnamefont {Trajkovic}},
  \bibinfo {author} {\bibfnamefont {I.}~\bibnamefont {O'Connor}}, \bibinfo
  {author} {\bibfnamefont {G.}~\bibnamefont {Nicolescu}}, \bibinfo {author}
  {\bibfnamefont {G.}~\bibnamefont {Bois}}, \ and\ \bibinfo {author}
  {\bibfnamefont {P.}~\bibnamefont {Paulin}},\ }in\ \href {\doibase
  10.1109/DATE.2011.5763134} {\emph {\bibinfo {booktitle} {Design, Automation
  Test in Europe Conference Exhibition (DATE), 2011}}}\ (\bibinfo {year}
  {2011})\ pp.\ \bibinfo {pages} {1--6}\BibitemShut {NoStop}%
\bibitem [{\citenamefont {Narayana}\ \emph {et~al.}(2017)\citenamefont
  {Narayana}, \citenamefont {Sun}, \citenamefont {Badawy}, \citenamefont
  {Sorger},\ and\ \citenamefont {El-Ghazawi}}]{Narayana:17}%
  \BibitemOpen
  \bibfield  {author} {\bibinfo {author} {\bibfnamefont {V.~K.}\ \bibnamefont
  {Narayana}}, \bibinfo {author} {\bibfnamefont {S.}~\bibnamefont {Sun}},
  \bibinfo {author} {\bibfnamefont {A.-H.~A.}\ \bibnamefont {Badawy}}, \bibinfo
  {author} {\bibfnamefont {V.~J.}\ \bibnamefont {Sorger}}, \ and\ \bibinfo
  {author} {\bibfnamefont {T.}~\bibnamefont {El-Ghazawi}},\ }\href@noop {}
  {\bibfield  {journal} {\bibinfo  {journal} {arXiv:1506.03264}\ } (\bibinfo
  {year} {2017})}\BibitemShut {NoStop}%
\bibitem [{\citenamefont {Capmany}\ \emph {et~al.}(2013)\citenamefont
  {Capmany}, \citenamefont {Mora}, \citenamefont {Gasulla}, \citenamefont
  {Sancho}, \citenamefont {Lloret},\ and\ \citenamefont {Sales}}]{Capmany:13}%
  \BibitemOpen
  \bibfield  {author} {\bibinfo {author} {\bibfnamefont {J.}~\bibnamefont
  {Capmany}}, \bibinfo {author} {\bibfnamefont {J.}~\bibnamefont {Mora}},
  \bibinfo {author} {\bibfnamefont {I.}~\bibnamefont {Gasulla}}, \bibinfo
  {author} {\bibfnamefont {J.}~\bibnamefont {Sancho}}, \bibinfo {author}
  {\bibfnamefont {J.}~\bibnamefont {Lloret}}, \ and\ \bibinfo {author}
  {\bibfnamefont {S.}~\bibnamefont {Sales}},\ }\href {\doibase
  10.1109/JLT.2012.2222348} {\bibfield  {journal} {\bibinfo  {journal} {Journal
  of Lightwave Technology}\ }\textbf {\bibinfo {volume} {31}},\ \bibinfo
  {pages} {571} (\bibinfo {year} {2013})}\BibitemShut {NoStop}%
\bibitem [{\citenamefont {Farsaei}\ \emph {et~al.}(2016)\citenamefont
  {Farsaei}, \citenamefont {Wang}, \citenamefont {Molavi}, \citenamefont
  {Jayatilleka}, \citenamefont {Caverley}, \citenamefont {Beikahmadi},
  \citenamefont {Shirazi}, \citenamefont {Jaeger}, \citenamefont
  {Chrostowski},\ and\ \citenamefont {Mirabbasi}}]{Farsaei:16}%
  \BibitemOpen
  \bibfield  {author} {\bibinfo {author} {\bibfnamefont {A.}~\bibnamefont
  {Farsaei}}, \bibinfo {author} {\bibfnamefont {Y.}~\bibnamefont {Wang}},
  \bibinfo {author} {\bibfnamefont {R.}~\bibnamefont {Molavi}}, \bibinfo
  {author} {\bibfnamefont {H.}~\bibnamefont {Jayatilleka}}, \bibinfo {author}
  {\bibfnamefont {M.}~\bibnamefont {Caverley}}, \bibinfo {author}
  {\bibfnamefont {M.}~\bibnamefont {Beikahmadi}}, \bibinfo {author}
  {\bibfnamefont {A.~H.~M.}\ \bibnamefont {Shirazi}}, \bibinfo {author}
  {\bibfnamefont {N.}~\bibnamefont {Jaeger}}, \bibinfo {author} {\bibfnamefont
  {L.}~\bibnamefont {Chrostowski}}, \ and\ \bibinfo {author} {\bibfnamefont
  {S.}~\bibnamefont {Mirabbasi}},\ }\href {\doibase
  http://dx.doi.org/10.1016/j.optcom.2016.01.074} {\bibfield  {journal}
  {\bibinfo  {journal} {Optics Communications}\ ,\ } (\bibinfo {year}
  {2016})}\BibitemShut {NoStop}%
\bibitem [{\citenamefont {Feng}\ \emph {et~al.}(2010)\citenamefont {Feng},
  \citenamefont {Dong}, \citenamefont {Feng}, \citenamefont {Qian},
  \citenamefont {Liang}, \citenamefont {Lee}, \citenamefont {Luff},
  \citenamefont {Agarwal}, \citenamefont {Banwell}, \citenamefont {Menendez},
  \citenamefont {Toliver}, \citenamefont {Woodward},\ and\ \citenamefont
  {Asghari}}]{Feng:10}%
  \BibitemOpen
  \bibfield  {author} {\bibinfo {author} {\bibfnamefont {N.-N.}\ \bibnamefont
  {Feng}}, \bibinfo {author} {\bibfnamefont {P.}~\bibnamefont {Dong}}, \bibinfo
  {author} {\bibfnamefont {D.}~\bibnamefont {Feng}}, \bibinfo {author}
  {\bibfnamefont {W.}~\bibnamefont {Qian}}, \bibinfo {author} {\bibfnamefont
  {H.}~\bibnamefont {Liang}}, \bibinfo {author} {\bibfnamefont {D.~C.}\
  \bibnamefont {Lee}}, \bibinfo {author} {\bibfnamefont {J.~B.}\ \bibnamefont
  {Luff}}, \bibinfo {author} {\bibfnamefont {A.}~\bibnamefont {Agarwal}},
  \bibinfo {author} {\bibfnamefont {T.}~\bibnamefont {Banwell}}, \bibinfo
  {author} {\bibfnamefont {R.}~\bibnamefont {Menendez}}, \bibinfo {author}
  {\bibfnamefont {P.}~\bibnamefont {Toliver}}, \bibinfo {author} {\bibfnamefont
  {T.~K.}\ \bibnamefont {Woodward}}, \ and\ \bibinfo {author} {\bibfnamefont
  {M.}~\bibnamefont {Asghari}},\ }\href {\doibase 10.1364/OE.18.024648}
  {\bibfield  {journal} {\bibinfo  {journal} {Opt. Express}\ }\textbf {\bibinfo
  {volume} {18}},\ \bibinfo {pages} {24648} (\bibinfo {year}
  {2010})}\BibitemShut {NoStop}%
\bibitem [{\citenamefont {Zhuang}\ \emph {et~al.}(2015)\citenamefont {Zhuang},
  \citenamefont {Roeloffzen}, \citenamefont {Hoekman}, \citenamefont {Boller},\
  and\ \citenamefont {Lowery}}]{Zhuang:15}%
  \BibitemOpen
  \bibfield  {author} {\bibinfo {author} {\bibfnamefont {L.}~\bibnamefont
  {Zhuang}}, \bibinfo {author} {\bibfnamefont {C.~G.~H.}\ \bibnamefont
  {Roeloffzen}}, \bibinfo {author} {\bibfnamefont {M.}~\bibnamefont {Hoekman}},
  \bibinfo {author} {\bibfnamefont {K.-J.}\ \bibnamefont {Boller}}, \ and\
  \bibinfo {author} {\bibfnamefont {A.~J.}\ \bibnamefont {Lowery}},\ }\href
  {\doibase 10.1364/OPTICA.2.000854} {\bibfield  {journal} {\bibinfo  {journal}
  {Optica}\ }\textbf {\bibinfo {volume} {2}},\ \bibinfo {pages} {854} (\bibinfo
  {year} {2015})}\BibitemShut {NoStop}%
\bibitem [{\citenamefont {Valley}(2007)}]{Valley:07}%
  \BibitemOpen
  \bibfield  {author} {\bibinfo {author} {\bibfnamefont {G.~C.}\ \bibnamefont
  {Valley}},\ }\href {\doibase 10.1364/OE.15.001955} {\bibfield  {journal}
  {\bibinfo  {journal} {Opt. Express}\ }\textbf {\bibinfo {volume} {15}},\
  \bibinfo {pages} {1955} (\bibinfo {year} {2007})}\BibitemShut {NoStop}%
\bibitem [{\citenamefont {Khan}\ \emph {et~al.}(2010)\citenamefont {Khan},
  \citenamefont {Shen}, \citenamefont {Xuan}, \citenamefont {Zhao},
  \citenamefont {Xiao}, \citenamefont {Leaird}, \citenamefont {Weiner},\ and\
  \citenamefont {Qi}}]{Khan:2010}%
  \BibitemOpen
  \bibfield  {author} {\bibinfo {author} {\bibfnamefont {M.~H.}\ \bibnamefont
  {Khan}}, \bibinfo {author} {\bibfnamefont {H.}~\bibnamefont {Shen}}, \bibinfo
  {author} {\bibfnamefont {Y.}~\bibnamefont {Xuan}}, \bibinfo {author}
  {\bibfnamefont {L.}~\bibnamefont {Zhao}}, \bibinfo {author} {\bibfnamefont
  {S.}~\bibnamefont {Xiao}}, \bibinfo {author} {\bibfnamefont {D.~E.}\
  \bibnamefont {Leaird}}, \bibinfo {author} {\bibfnamefont {A.~M.}\
  \bibnamefont {Weiner}}, \ and\ \bibinfo {author} {\bibfnamefont
  {M.}~\bibnamefont {Qi}},\ }\href {\doibase 10.1038/nphoton.2009.266}
  {\bibfield  {journal} {\bibinfo  {journal} {Nature: Photonics}\ }\textbf
  {\bibinfo {volume} {4}},\ \bibinfo {pages} {117} (\bibinfo {year}
  {2010})}\BibitemShut {NoStop}%
\bibitem [{\citenamefont {Chang}\ \emph {et~al.}(2014)\citenamefont {Chang},
  \citenamefont {Meister},\ and\ \citenamefont {Prucnal}}]{Chang:14implement}%
  \BibitemOpen
  \bibfield  {author} {\bibinfo {author} {\bibfnamefont {J.}~\bibnamefont
  {Chang}}, \bibinfo {author} {\bibfnamefont {J.}~\bibnamefont {Meister}}, \
  and\ \bibinfo {author} {\bibfnamefont {P.~R.}\ \bibnamefont {Prucnal}},\
  }\href {\doibase 10.1109/JLT.2014.2309691} {\bibfield  {journal} {\bibinfo
  {journal} {Journal of Lightwave Technology}\ }\textbf {\bibinfo {volume}
  {32}},\ \bibinfo {pages} {3623} (\bibinfo {year} {2014})}\BibitemShut
  {NoStop}%
\bibitem [{\citenamefont {{Ferreira de Lima}}\ \emph
  {et~al.}(2016)\citenamefont {{Ferreira de Lima}}, \citenamefont {Tait},
  \citenamefont {Nahmias}, \citenamefont {Shastri},\ and\ \citenamefont
  {Prucnal}}]{FerreiradeLima:16}%
  \BibitemOpen
  \bibfield  {author} {\bibinfo {author} {\bibfnamefont {T.}~\bibnamefont
  {{Ferreira de Lima}}}, \bibinfo {author} {\bibfnamefont {A.~N.}\ \bibnamefont
  {Tait}}, \bibinfo {author} {\bibfnamefont {M.~A.}\ \bibnamefont {Nahmias}},
  \bibinfo {author} {\bibfnamefont {B.~J.}\ \bibnamefont {Shastri}}, \ and\
  \bibinfo {author} {\bibfnamefont {P.~R.}\ \bibnamefont {Prucnal}},\ }\href
  {\doibase 10.1109/JPHOT.2016.2538759} {\bibfield  {journal} {\bibinfo
  {journal} {IEEE Photonics Journal}\ }\textbf {\bibinfo {volume} {8}},\
  \bibinfo {pages} {1} (\bibinfo {year} {2016})}\BibitemShut {NoStop}%
\bibitem [{\citenamefont {Merolla}\ \emph {et~al.}(2014)\citenamefont
  {Merolla}, \citenamefont {Arthur}, \citenamefont {Alvarez-Icaza},
  \citenamefont {Cassidy}, \citenamefont {Sawada}, \citenamefont {Akopyan},
  \citenamefont {Jackson}, \citenamefont {Imam}, \citenamefont {Guo},
  \citenamefont {Nakamura}, \citenamefont {Brezzo}, \citenamefont {Vo},
  \citenamefont {Esser}, \citenamefont {Appuswamy}, \citenamefont {Taba},
  \citenamefont {Amir}, \citenamefont {Flickner}, \citenamefont {Risk},
  \citenamefont {Manohar},\ and\ \citenamefont {Modha}}]{Merolla:2014}%
  \BibitemOpen
  \bibfield  {author} {\bibinfo {author} {\bibfnamefont {P.~A.}\ \bibnamefont
  {Merolla}}, \bibinfo {author} {\bibfnamefont {J.~V.}\ \bibnamefont {Arthur}},
  \bibinfo {author} {\bibfnamefont {R.}~\bibnamefont {Alvarez-Icaza}}, \bibinfo
  {author} {\bibfnamefont {A.~S.}\ \bibnamefont {Cassidy}}, \bibinfo {author}
  {\bibfnamefont {J.}~\bibnamefont {Sawada}}, \bibinfo {author} {\bibfnamefont
  {F.}~\bibnamefont {Akopyan}}, \bibinfo {author} {\bibfnamefont {B.~L.}\
  \bibnamefont {Jackson}}, \bibinfo {author} {\bibfnamefont {N.}~\bibnamefont
  {Imam}}, \bibinfo {author} {\bibfnamefont {C.}~\bibnamefont {Guo}}, \bibinfo
  {author} {\bibfnamefont {Y.}~\bibnamefont {Nakamura}}, \bibinfo {author}
  {\bibfnamefont {B.}~\bibnamefont {Brezzo}}, \bibinfo {author} {\bibfnamefont
  {I.}~\bibnamefont {Vo}}, \bibinfo {author} {\bibfnamefont {S.~K.}\
  \bibnamefont {Esser}}, \bibinfo {author} {\bibfnamefont {R.}~\bibnamefont
  {Appuswamy}}, \bibinfo {author} {\bibfnamefont {B.}~\bibnamefont {Taba}},
  \bibinfo {author} {\bibfnamefont {A.}~\bibnamefont {Amir}}, \bibinfo {author}
  {\bibfnamefont {M.~D.}\ \bibnamefont {Flickner}}, \bibinfo {author}
  {\bibfnamefont {W.~P.}\ \bibnamefont {Risk}}, \bibinfo {author}
  {\bibfnamefont {R.}~\bibnamefont {Manohar}}, \ and\ \bibinfo {author}
  {\bibfnamefont {D.~S.}\ \bibnamefont {Modha}},\ }\href {\doibase
  10.1126/science.1254642} {\bibfield  {journal} {\bibinfo  {journal}
  {Science}\ }\textbf {\bibinfo {volume} {345}},\ \bibinfo {pages} {668}
  (\bibinfo {year} {2014})}\BibitemShut {NoStop}%
\bibitem [{\citenamefont {Akopyan}\ \emph {et~al.}(2015)\citenamefont
  {Akopyan}, \citenamefont {Sawada}, \citenamefont {Cassidy}, \citenamefont
  {Alvarez-Icaza}, \citenamefont {Arthur}, \citenamefont {Merolla},
  \citenamefont {Imam}, \citenamefont {Nakamura}, \citenamefont {Datta},
  \citenamefont {Nam}, \citenamefont {Taba}, \citenamefont {Beakes},
  \citenamefont {Brezzo}, \citenamefont {Kuang}, \citenamefont {Manohar},
  \citenamefont {Risk}, \citenamefont {Jackson},\ and\ \citenamefont
  {Modha}}]{Akopyan:15}%
  \BibitemOpen
  \bibfield  {author} {\bibinfo {author} {\bibfnamefont {F.}~\bibnamefont
  {Akopyan}}, \bibinfo {author} {\bibfnamefont {J.}~\bibnamefont {Sawada}},
  \bibinfo {author} {\bibfnamefont {A.}~\bibnamefont {Cassidy}}, \bibinfo
  {author} {\bibfnamefont {R.}~\bibnamefont {Alvarez-Icaza}}, \bibinfo {author}
  {\bibfnamefont {J.}~\bibnamefont {Arthur}}, \bibinfo {author} {\bibfnamefont
  {P.}~\bibnamefont {Merolla}}, \bibinfo {author} {\bibfnamefont
  {N.}~\bibnamefont {Imam}}, \bibinfo {author} {\bibfnamefont {Y.}~\bibnamefont
  {Nakamura}}, \bibinfo {author} {\bibfnamefont {P.}~\bibnamefont {Datta}},
  \bibinfo {author} {\bibfnamefont {G.-J.}\ \bibnamefont {Nam}}, \bibinfo
  {author} {\bibfnamefont {B.}~\bibnamefont {Taba}}, \bibinfo {author}
  {\bibfnamefont {M.}~\bibnamefont {Beakes}}, \bibinfo {author} {\bibfnamefont
  {B.}~\bibnamefont {Brezzo}}, \bibinfo {author} {\bibfnamefont
  {J.}~\bibnamefont {Kuang}}, \bibinfo {author} {\bibfnamefont
  {R.}~\bibnamefont {Manohar}}, \bibinfo {author} {\bibfnamefont
  {W.}~\bibnamefont {Risk}}, \bibinfo {author} {\bibfnamefont {B.}~\bibnamefont
  {Jackson}}, \ and\ \bibinfo {author} {\bibfnamefont {D.}~\bibnamefont
  {Modha}},\ }\href {\doibase 10.1109/TCAD.2015.2474396} {\bibfield  {journal}
  {\bibinfo  {journal} {IEEE Trans. Comput. Aided Des. Integr. Circuits Syst.}\
  }\textbf {\bibinfo {volume} {34}},\ \bibinfo {pages} {1537} (\bibinfo {year}
  {2015})}\BibitemShut {NoStop}%
\bibitem [{\citenamefont {Indiveri}\ and\ \citenamefont
  {Liu}(2015)}]{Indiveri:15}%
  \BibitemOpen
  \bibfield  {author} {\bibinfo {author} {\bibfnamefont {G.}~\bibnamefont
  {Indiveri}}\ and\ \bibinfo {author} {\bibfnamefont {S.~C.}\ \bibnamefont
  {Liu}},\ }\href {\doibase 10.1109/JPROC.2015.2444094} {\bibfield  {journal}
  {\bibinfo  {journal} {Proceedings of the IEEE}\ }\textbf {\bibinfo {volume}
  {103}},\ \bibinfo {pages} {1379} (\bibinfo {year} {2015})}\BibitemShut
  {NoStop}%
\bibitem [{\citenamefont {Hasler}\ and\ \citenamefont
  {Marr}(2013)}]{Hasler2013}%
  \BibitemOpen
  \bibfield  {author} {\bibinfo {author} {\bibfnamefont {J.}~\bibnamefont
  {Hasler}}\ and\ \bibinfo {author} {\bibfnamefont {H.~B.}\ \bibnamefont
  {Marr}},\ }\href {\doibase 10.3389/fnins.2013.00118} {\bibfield  {journal}
  {\bibinfo  {journal} {Front. Neurosci.}\ }\textbf {\bibinfo {volume} {7}}
  (\bibinfo {year} {2013}),\ 10.3389/fnins.2013.00118}\BibitemShut {NoStop}%
\bibitem [{\citenamefont {Wen}\ \emph {et~al.}(2009)\citenamefont {Wen},
  \citenamefont {Lan},\ and\ \citenamefont {Shih}}]{Wen:09}%
  \BibitemOpen
  \bibfield  {author} {\bibinfo {author} {\bibfnamefont {U.-P.}\ \bibnamefont
  {Wen}}, \bibinfo {author} {\bibfnamefont {K.-M.}\ \bibnamefont {Lan}}, \ and\
  \bibinfo {author} {\bibfnamefont {H.-S.}\ \bibnamefont {Shih}},\ }\href
  {\doibase http://dx.doi.org/10.1016/j.ejor.2008.11.002} {\bibfield  {journal}
  {\bibinfo  {journal} {European Journal of Operational Research}\ }\textbf
  {\bibinfo {volume} {198}},\ \bibinfo {pages} {675 } (\bibinfo {year}
  {2009})}\BibitemShut {NoStop}%
\bibitem [{\citenamefont {Lee}\ and\ \citenamefont
  {Theunissen}(2015)}]{Lee:15}%
  \BibitemOpen
  \bibfield  {author} {\bibinfo {author} {\bibfnamefont {T.}~\bibnamefont
  {Lee}}\ and\ \bibinfo {author} {\bibfnamefont {F.}~\bibnamefont
  {Theunissen}},\ }\href {\doibase 10.1098/rspa.2015.0309} {\bibfield
  {journal} {\bibinfo  {journal} {Proceedings of the Royal Society of London A:
  Mathematical, Physical and Engineering Sciences}\ }\textbf {\bibinfo {volume}
  {471}} (\bibinfo {year} {2015}),\ 10.1098/rspa.2015.0309},\ \Eprint
  {http://arxiv.org/abs/http://rspa.royalsocietypublishing.org/content/471/2184/20150309.full.pdf}
  {http://rspa.royalsocietypublishing.org/content/471/2184/20150309.full.pdf}
  \BibitemShut {NoStop}%
\bibitem [{\citenamefont {Eliasmith}\ and\ \citenamefont
  {Anderson}(2004)}]{Eliasmith:04}%
  \BibitemOpen
  \bibfield  {author} {\bibinfo {author} {\bibfnamefont {C.}~\bibnamefont
  {Eliasmith}}\ and\ \bibinfo {author} {\bibfnamefont {C.~H.}\ \bibnamefont
  {Anderson}},\ }\href@noop {} {\emph {\bibinfo {title} {Neural engineering:
  Computation, representation, and dynamics in neurobiological systems}}}\
  (\bibinfo  {publisher} {MIT Press},\ \bibinfo {year} {2004})\BibitemShut
  {NoStop}%
\bibitem [{\citenamefont {Donnarumma}\ \emph {et~al.}(2016)\citenamefont
  {Donnarumma}, \citenamefont {Prevete}, \citenamefont {de~Giorgio},
  \citenamefont {Montone},\ and\ \citenamefont {Pezzulo}}]{Donnarumma:16}%
  \BibitemOpen
  \bibfield  {author} {\bibinfo {author} {\bibfnamefont {F.}~\bibnamefont
  {Donnarumma}}, \bibinfo {author} {\bibfnamefont {R.}~\bibnamefont {Prevete}},
  \bibinfo {author} {\bibfnamefont {A.}~\bibnamefont {de~Giorgio}}, \bibinfo
  {author} {\bibfnamefont {G.}~\bibnamefont {Montone}}, \ and\ \bibinfo
  {author} {\bibfnamefont {G.}~\bibnamefont {Pezzulo}},\ }\href {\doibase
  10.1177/1059712315609412} {\bibfield  {journal} {\bibinfo  {journal}
  {Adaptive Behavior}\ }\textbf {\bibinfo {volume} {24}},\ \bibinfo {pages}
  {27} (\bibinfo {year} {2016})},\ \Eprint
  {http://arxiv.org/abs/http://adb.sagepub.com/content/24/1/27.full.pdf+html}
  {http://adb.sagepub.com/content/24/1/27.full.pdf+html} \BibitemShut {NoStop}%
\bibitem [{\citenamefont {Diamond}\ \emph {et~al.}(2016)\citenamefont
  {Diamond}, \citenamefont {Nowotny},\ and\ \citenamefont
  {Schmuker}}]{Diamond:16}%
  \BibitemOpen
  \bibfield  {author} {\bibinfo {author} {\bibfnamefont {A.}~\bibnamefont
  {Diamond}}, \bibinfo {author} {\bibfnamefont {T.}~\bibnamefont {Nowotny}}, \
  and\ \bibinfo {author} {\bibfnamefont {M.}~\bibnamefont {Schmuker}},\ }\href
  {\doibase 10.3389/fnins.2015.00491} {\bibfield  {journal} {\bibinfo
  {journal} {Frontiers in Neuroscience}\ }\textbf {\bibinfo {volume} {9}}
  (\bibinfo {year} {2016}),\ 10.3389/fnins.2015.00491}\BibitemShut {NoStop}%
\bibitem [{\citenamefont {Tait}\ \emph {et~al.}(2014)\citenamefont {Tait},
  \citenamefont {Nahmias}, \citenamefont {Shastri},\ and\ \citenamefont
  {Prucnal}}]{Tait:14}%
  \BibitemOpen
  \bibfield  {author} {\bibinfo {author} {\bibfnamefont {A.~N.}\ \bibnamefont
  {Tait}}, \bibinfo {author} {\bibfnamefont {M.~A.}\ \bibnamefont {Nahmias}},
  \bibinfo {author} {\bibfnamefont {B.~J.}\ \bibnamefont {Shastri}}, \ and\
  \bibinfo {author} {\bibfnamefont {P.~R.}\ \bibnamefont {Prucnal}},\ }\href
  {\doibase 10.1109/JLT.2014.2345652} {\bibfield  {journal} {\bibinfo
  {journal} {Journal of Lightwave Technology}\ }\textbf {\bibinfo {volume}
  {32}},\ \bibinfo {pages} {4029} (\bibinfo {year} {2014})}\BibitemShut
  {NoStop}%
\bibitem [{\citenamefont {Brunner}\ and\ \citenamefont
  {Fischer}(2015)}]{Brunner:15}%
  \BibitemOpen
  \bibfield  {author} {\bibinfo {author} {\bibfnamefont {D.}~\bibnamefont
  {Brunner}}\ and\ \bibinfo {author} {\bibfnamefont {I.}~\bibnamefont
  {Fischer}},\ }\href {\doibase 10.1364/OL.40.003854} {\bibfield  {journal}
  {\bibinfo  {journal} {Optics letters}\ }\textbf {\bibinfo {volume} {40}},\
  \bibinfo {pages} {3854} (\bibinfo {year} {2015})}\BibitemShut {NoStop}%
\bibitem [{\citenamefont {Tait}\ \emph
  {et~al.}(2016{\natexlab{a}})\citenamefont {Tait}, \citenamefont {Wu},
  \citenamefont {{Ferreira de Lima}}, \citenamefont {Zhou}, \citenamefont
  {Shastri}, \citenamefont {Nahmias},\ and\ \citenamefont
  {Prucnal}}]{Tait:16scale}%
  \BibitemOpen
  \bibfield  {author} {\bibinfo {author} {\bibfnamefont {A.~N.}\ \bibnamefont
  {Tait}}, \bibinfo {author} {\bibfnamefont {A.~X.}\ \bibnamefont {Wu}},
  \bibinfo {author} {\bibfnamefont {T.}~\bibnamefont {{Ferreira de Lima}}},
  \bibinfo {author} {\bibfnamefont {E.}~\bibnamefont {Zhou}}, \bibinfo {author}
  {\bibfnamefont {B.~J.}\ \bibnamefont {Shastri}}, \bibinfo {author}
  {\bibfnamefont {M.~A.}\ \bibnamefont {Nahmias}}, \ and\ \bibinfo {author}
  {\bibfnamefont {P.~R.}\ \bibnamefont {Prucnal}},\ }\href {\doibase
  10.1109/JSTQE.2016.2573583} {\bibfield  {journal} {\bibinfo  {journal} {IEEE
  Journal of Selected Topics in Quantum Electronics}\ }\textbf {\bibinfo
  {volume} {22}} (\bibinfo {year} {2016}{\natexlab{a}}),\
  10.1109/JSTQE.2016.2573583}\BibitemShut {NoStop}%
\bibitem [{\citenamefont {Tait}\ \emph
  {et~al.}(2016{\natexlab{b}})\citenamefont {Tait}, \citenamefont {{Ferreira de
  Lima}}, \citenamefont {Nahmias}, \citenamefont {Shastri},\ and\ \citenamefont
  {Prucnal}}]{Tait:16multi}%
  \BibitemOpen
  \bibfield  {author} {\bibinfo {author} {\bibfnamefont {A.~N.}\ \bibnamefont
  {Tait}}, \bibinfo {author} {\bibfnamefont {T.}~\bibnamefont {{Ferreira de
  Lima}}}, \bibinfo {author} {\bibfnamefont {M.~A.}\ \bibnamefont {Nahmias}},
  \bibinfo {author} {\bibfnamefont {B.~J.}\ \bibnamefont {Shastri}}, \ and\
  \bibinfo {author} {\bibfnamefont {P.~R.}\ \bibnamefont {Prucnal}},\ }\href
  {\doibase 10.1364/OE.24.008895} {\bibfield  {journal} {\bibinfo  {journal}
  {Opt. Express}\ }\textbf {\bibinfo {volume} {24}},\ \bibinfo {pages} {8895}
  (\bibinfo {year} {2016}{\natexlab{b}})}\BibitemShut {NoStop}%
\bibitem [{\citenamefont {Yamada}(1993)}]{Yamada:93}%
  \BibitemOpen
  \bibfield  {author} {\bibinfo {author} {\bibfnamefont {M.}~\bibnamefont
  {Yamada}},\ }\href {\doibase 10.1109/3.236146} {\bibfield  {journal}
  {\bibinfo  {journal} {IEEE Journal of Quantum Electronics}\ }\textbf
  {\bibinfo {volume} {29}},\ \bibinfo {pages} {1330} (\bibinfo {year}
  {1993})}\BibitemShut {NoStop}%
\bibitem [{\citenamefont {Romeira}\ \emph {et~al.}(2014)\citenamefont
  {Romeira}, \citenamefont {Kong}, \citenamefont {Li}, \citenamefont
  {Figueiredo}, \citenamefont {Javaloyes},\ and\ \citenamefont
  {Yao}}]{Romeira:14}%
  \BibitemOpen
  \bibfield  {author} {\bibinfo {author} {\bibfnamefont {B.}~\bibnamefont
  {Romeira}}, \bibinfo {author} {\bibfnamefont {F.}~\bibnamefont {Kong}},
  \bibinfo {author} {\bibfnamefont {W.}~\bibnamefont {Li}}, \bibinfo {author}
  {\bibfnamefont {J.~M.}\ \bibnamefont {Figueiredo}}, \bibinfo {author}
  {\bibfnamefont {J.}~\bibnamefont {Javaloyes}}, \ and\ \bibinfo {author}
  {\bibfnamefont {J.}~\bibnamefont {Yao}},\ }\href {\doibase
  10.1109/JLT.2014.2308261} {\bibfield  {journal} {\bibinfo  {journal}
  {Lightwave Technology, Journal of}\ }\textbf {\bibinfo {volume} {32}},\
  \bibinfo {pages} {3933} (\bibinfo {year} {2014})}\BibitemShut {NoStop}%
\bibitem [{\citenamefont {Beer}(1995)}]{Beer:95}%
  \BibitemOpen
  \bibfield  {author} {\bibinfo {author} {\bibfnamefont {R.~D.}\ \bibnamefont
  {Beer}},\ }\href {\doibase 10.1177/105971239500300405} {\bibfield  {journal}
  {\bibinfo  {journal} {Adaptive Behavior}\ }\textbf {\bibinfo {volume} {3}},\
  \bibinfo {pages} {469} (\bibinfo {year} {1995})}\BibitemShut {NoStop}%
\bibitem [{\citenamefont {Zhou}\ \emph {et~al.}(2016)\citenamefont {Zhou},
  \citenamefont {Tait}, \citenamefont {Wu}, \citenamefont {{Ferreira de Lima}},
  \citenamefont {Nahmias}, \citenamefont {Shastri},\ and\ \citenamefont
  {Prucnal}}]{Zhou:16oi}%
  \BibitemOpen
  \bibfield  {author} {\bibinfo {author} {\bibfnamefont {E.}~\bibnamefont
  {Zhou}}, \bibinfo {author} {\bibfnamefont {A.}~\bibnamefont {Tait}}, \bibinfo
  {author} {\bibfnamefont {A.}~\bibnamefont {Wu}}, \bibinfo {author}
  {\bibfnamefont {T.}~\bibnamefont {{Ferreira de Lima}}}, \bibinfo {author}
  {\bibfnamefont {M.}~\bibnamefont {Nahmias}}, \bibinfo {author} {\bibfnamefont
  {B.}~\bibnamefont {Shastri}}, \ and\ \bibinfo {author} {\bibfnamefont
  {P.}~\bibnamefont {Prucnal}},\ }in\ \href {\doibase 10.1109/OIC.2016.7483010}
  {\emph {\bibinfo {booktitle} {Optical Interconnects Conference, 2016 IEEE}}}\
  (\bibinfo  {publisher} {IEEE},\ \bibinfo {year} {2016})\ p.\ \bibinfo {pages}
  {TuP9}\BibitemShut {NoStop}%
\bibitem [{\citenamefont {Stewart}\ and\ \citenamefont
  {Eliasmith}(2014)}]{stewart2014large}%
  \BibitemOpen
  \bibfield  {author} {\bibinfo {author} {\bibfnamefont {T.~C.}\ \bibnamefont
  {Stewart}}\ and\ \bibinfo {author} {\bibfnamefont {C.}~\bibnamefont
  {Eliasmith}},\ }\href {\doibase 10.1109/JPROC.2014.2306061} {\bibfield
  {journal} {\bibinfo  {journal} {Proceedings of the IEEE}\ }\textbf {\bibinfo
  {volume} {102}},\ \bibinfo {pages} {881} (\bibinfo {year}
  {2014})}\BibitemShut {NoStop}%
\bibitem [{\citenamefont {Khanna}(2015)}]{Khanna:15}%
  \BibitemOpen
  \bibfield  {author} {\bibinfo {author} {\bibfnamefont {A.}~\bibnamefont
  {Khanna}},\ }in\ \href@noop {} {\emph {\bibinfo {booktitle} {European
  Conference on Optical Communication}}}\ (\bibinfo {year} {2015})\BibitemShut
  {NoStop}%
\bibitem [{\citenamefont {Roska}\ \emph {et~al.}(1995)\citenamefont {Roska},
  \citenamefont {Chua}, \citenamefont {Wolf}, \citenamefont {Kozek},
  \citenamefont {Tetzlaff},\ and\ \citenamefont {Puffer}}]{Roska:95}%
  \BibitemOpen
  \bibfield  {author} {\bibinfo {author} {\bibfnamefont {T.}~\bibnamefont
  {Roska}}, \bibinfo {author} {\bibfnamefont {L.}~\bibnamefont {Chua}},
  \bibinfo {author} {\bibfnamefont {D.}~\bibnamefont {Wolf}}, \bibinfo {author}
  {\bibfnamefont {T.}~\bibnamefont {Kozek}}, \bibinfo {author} {\bibfnamefont
  {R.}~\bibnamefont {Tetzlaff}}, \ and\ \bibinfo {author} {\bibfnamefont
  {F.}~\bibnamefont {Puffer}},\ }\href {\doibase 10.1109/81.473590} {\bibfield
  {journal} {\bibinfo  {journal} {Circuits and Systems I: Fundamental Theory
  and Applications, IEEE Transactions on}\ }\textbf {\bibinfo {volume} {42}},\
  \bibinfo {pages} {807} (\bibinfo {year} {1995})}\BibitemShut {NoStop}%
\bibitem [{\citenamefont {Ratier}(2012)}]{Ratier:12}%
  \BibitemOpen
  \bibfield  {author} {\bibinfo {author} {\bibfnamefont {N.}~\bibnamefont
  {Ratier}},\ }in\ \href {\doibase 10.1109/SETIT.2012.6481928} {\emph {\bibinfo
  {booktitle} {Sciences of Electronics, Technologies of Information and
  Telecommunications (SETIT), 2012 6th International Conference on}}}\
  (\bibinfo {year} {2012})\ pp.\ \bibinfo {pages} {275--282}\BibitemShut
  {NoStop}%
\bibitem [{\citenamefont {Vogelstein}\ \emph {et~al.}(2008)\citenamefont
  {Vogelstein}, \citenamefont {Tenore}, \citenamefont {Guevremont},
  \citenamefont {Etienne-Cummings},\ and\ \citenamefont
  {Mushahwar}}]{Vogelstein:08}%
  \BibitemOpen
  \bibfield  {author} {\bibinfo {author} {\bibfnamefont {R.~J.}\ \bibnamefont
  {Vogelstein}}, \bibinfo {author} {\bibfnamefont {F.~V.~G.}\ \bibnamefont
  {Tenore}}, \bibinfo {author} {\bibfnamefont {L.}~\bibnamefont {Guevremont}},
  \bibinfo {author} {\bibfnamefont {R.}~\bibnamefont {Etienne-Cummings}}, \
  and\ \bibinfo {author} {\bibfnamefont {V.~K.}\ \bibnamefont {Mushahwar}},\
  }\href {\doibase 10.1109/TBCAS.2008.2001867} {\bibfield  {journal} {\bibinfo
  {journal} {IEEE Transactions on Biomedical Circuits and Systems}\ }\textbf
  {\bibinfo {volume} {2}},\ \bibinfo {pages} {212} (\bibinfo {year}
  {2008})}\BibitemShut {NoStop}%
\bibitem [{\citenamefont {Arena}\ \emph {et~al.}(2005)\citenamefont {Arena},
  \citenamefont {Fortuna}, \citenamefont {Frasca},\ and\ \citenamefont
  {Patane}}]{Arena:05}%
  \BibitemOpen
  \bibfield  {author} {\bibinfo {author} {\bibfnamefont {P.}~\bibnamefont
  {Arena}}, \bibinfo {author} {\bibfnamefont {L.}~\bibnamefont {Fortuna}},
  \bibinfo {author} {\bibfnamefont {M.}~\bibnamefont {Frasca}}, \ and\ \bibinfo
  {author} {\bibfnamefont {L.}~\bibnamefont {Patane}},\ }\href {\doibase
  10.1109/TCSI.2005.852211} {\bibfield  {journal} {\bibinfo  {journal} {IEEE
  Transactions on Circuits and Systems I: Regular Papers}\ }\textbf {\bibinfo
  {volume} {52}},\ \bibinfo {pages} {1862} (\bibinfo {year}
  {2005})}\BibitemShut {NoStop}%
\bibitem [{\citenamefont {Barron-Zambrano}\ and\ \citenamefont
  {Torres-Huitzil}(2013)}]{BarronZambrano:13}%
  \BibitemOpen
  \bibfield  {author} {\bibinfo {author} {\bibfnamefont {J.~H.}\ \bibnamefont
  {Barron-Zambrano}}\ and\ \bibinfo {author} {\bibfnamefont {C.}~\bibnamefont
  {Torres-Huitzil}},\ }\href {\doibase
  http://doi.org/10.1016/j.neunet.2013.04.005} {\bibfield  {journal} {\bibinfo
  {journal} {Neural Networks}\ }\textbf {\bibinfo {volume} {45}},\ \bibinfo
  {pages} {50 } (\bibinfo {year} {2013})},\ \bibinfo {note} {neuromorphic
  Engineering: From Neural Systems to Brain-Like Engineered
  Systems}\BibitemShut {NoStop}%
\bibitem [{\citenamefont {Friedmann}\ \emph {et~al.}(2013)\citenamefont
  {Friedmann}, \citenamefont {Fr\'emaux}, \citenamefont {Schemmel},
  \citenamefont {Gerstner},\ and\ \citenamefont {Meier}}]{Friedmann:13}%
  \BibitemOpen
  \bibfield  {author} {\bibinfo {author} {\bibfnamefont {S.}~\bibnamefont
  {Friedmann}}, \bibinfo {author} {\bibfnamefont {N.}~\bibnamefont
  {Fr\'emaux}}, \bibinfo {author} {\bibfnamefont {J.}~\bibnamefont {Schemmel}},
  \bibinfo {author} {\bibfnamefont {W.}~\bibnamefont {Gerstner}}, \ and\
  \bibinfo {author} {\bibfnamefont {K.}~\bibnamefont {Meier}},\ }\href
  {\doibase 10.3389/fnins.2013.00160} {\bibfield  {journal} {\bibinfo
  {journal} {Front. Neurosci.}\ }\textbf {\bibinfo {volume} {7}} (\bibinfo
  {year} {2013}),\ 10.3389/fnins.2013.00160}\BibitemShut {NoStop}%
\bibitem [{\citenamefont {Benjamin}\ \emph {et~al.}(2014)\citenamefont
  {Benjamin}, \citenamefont {Gao}, \citenamefont {McQuinn}, \citenamefont
  {Choudhary}, \citenamefont {Chandrasekaran}, \citenamefont {Bussat},
  \citenamefont {Alvarez-Icaza}, \citenamefont {Arthur}, \citenamefont
  {Merolla},\ and\ \citenamefont {Boahen}}]{Benjamin:14}%
  \BibitemOpen
  \bibfield  {author} {\bibinfo {author} {\bibfnamefont {B.}~\bibnamefont
  {Benjamin}}, \bibinfo {author} {\bibfnamefont {P.}~\bibnamefont {Gao}},
  \bibinfo {author} {\bibfnamefont {E.}~\bibnamefont {McQuinn}}, \bibinfo
  {author} {\bibfnamefont {S.}~\bibnamefont {Choudhary}}, \bibinfo {author}
  {\bibfnamefont {A.}~\bibnamefont {Chandrasekaran}}, \bibinfo {author}
  {\bibfnamefont {J.-M.}\ \bibnamefont {Bussat}}, \bibinfo {author}
  {\bibfnamefont {R.}~\bibnamefont {Alvarez-Icaza}}, \bibinfo {author}
  {\bibfnamefont {J.}~\bibnamefont {Arthur}}, \bibinfo {author} {\bibfnamefont
  {P.}~\bibnamefont {Merolla}}, \ and\ \bibinfo {author} {\bibfnamefont
  {K.}~\bibnamefont {Boahen}},\ }\href {\doibase 10.1109/JPROC.2014.2313565}
  {\bibfield  {journal} {\bibinfo  {journal} {Proceedings of the IEEE}\
  }\textbf {\bibinfo {volume} {102}},\ \bibinfo {pages} {699} (\bibinfo {year}
  {2014})}\BibitemShut {NoStop}%
\bibitem [{\citenamefont {Pfeil}\ \emph {et~al.}(2012)\citenamefont {Pfeil},
  \citenamefont {Potjans}, \citenamefont {Schrader}, \citenamefont {Potjans},
  \citenamefont {Schemmel}, \citenamefont {Diesmann},\ and\ \citenamefont
  {Meier}}]{Pfeil:12}%
  \BibitemOpen
  \bibfield  {author} {\bibinfo {author} {\bibfnamefont {T.}~\bibnamefont
  {Pfeil}}, \bibinfo {author} {\bibfnamefont {T.}~\bibnamefont {Potjans}},
  \bibinfo {author} {\bibfnamefont {S.}~\bibnamefont {Schrader}}, \bibinfo
  {author} {\bibfnamefont {W.}~\bibnamefont {Potjans}}, \bibinfo {author}
  {\bibfnamefont {J.}~\bibnamefont {Schemmel}}, \bibinfo {author}
  {\bibfnamefont {M.}~\bibnamefont {Diesmann}}, \ and\ \bibinfo {author}
  {\bibfnamefont {K.}~\bibnamefont {Meier}},\ }\href {\doibase
  10.3389/fnins.2012.00090} {\bibfield  {journal} {\bibinfo  {journal}
  {Frontiers in Neuroscience}\ }\textbf {\bibinfo {volume} {6}},\ \bibinfo
  {pages} {90} (\bibinfo {year} {2012})}\BibitemShut {NoStop}%
\bibitem [{\citenamefont {Binas}\ \emph {et~al.}(2016)\citenamefont {Binas},
  \citenamefont {Neil}, \citenamefont {Indiveri}, \citenamefont {Liu},\ and\
  \citenamefont {Pfeiffer}}]{Binas:16}%
  \BibitemOpen
  \bibfield  {author} {\bibinfo {author} {\bibfnamefont {J.}~\bibnamefont
  {Binas}}, \bibinfo {author} {\bibfnamefont {D.}~\bibnamefont {Neil}},
  \bibinfo {author} {\bibfnamefont {G.}~\bibnamefont {Indiveri}}, \bibinfo
  {author} {\bibfnamefont {S.-C.}\ \bibnamefont {Liu}}, \ and\ \bibinfo
  {author} {\bibfnamefont {M.}~\bibnamefont {Pfeiffer}},\ }\href@noop {}
  {\bibfield  {journal} {\bibinfo  {journal} {arXiv preprint arXiv:1606.07786}\
  } (\bibinfo {year} {2016})}\BibitemShut {NoStop}%
\bibitem [{\citenamefont {Shen}\ \emph {et~al.}(2016)\citenamefont {Shen},
  \citenamefont {Harris}, \citenamefont {Skirlo}, \citenamefont {Prabhu},
  \citenamefont {Baehr-Jones}, \citenamefont {Hochberg}, \citenamefont {Sun},
  \citenamefont {Zhao}, \citenamefont {Larochelle}, \citenamefont {Englund},\
  and\ \citenamefont {Soljacic}}]{Shen:16arxiv}%
  \BibitemOpen
  \bibfield  {author} {\bibinfo {author} {\bibfnamefont {Y.}~\bibnamefont
  {Shen}}, \bibinfo {author} {\bibfnamefont {N.~C.}\ \bibnamefont {Harris}},
  \bibinfo {author} {\bibfnamefont {S.}~\bibnamefont {Skirlo}}, \bibinfo
  {author} {\bibfnamefont {M.}~\bibnamefont {Prabhu}}, \bibinfo {author}
  {\bibfnamefont {T.}~\bibnamefont {Baehr-Jones}}, \bibinfo {author}
  {\bibfnamefont {M.}~\bibnamefont {Hochberg}}, \bibinfo {author}
  {\bibfnamefont {X.}~\bibnamefont {Sun}}, \bibinfo {author} {\bibfnamefont
  {S.}~\bibnamefont {Zhao}}, \bibinfo {author} {\bibfnamefont {H.}~\bibnamefont
  {Larochelle}}, \bibinfo {author} {\bibfnamefont {D.}~\bibnamefont {Englund}},
  \ and\ \bibinfo {author} {\bibfnamefont {M.}~\bibnamefont {Soljacic}},\
  }\href@noop {} {\bibfield  {journal} {\bibinfo  {journal} {arXiv:1610.02365}\
  } (\bibinfo {year} {2016})}\BibitemShut {NoStop}%
\bibitem [{\citenamefont {Shainline}\ \emph {et~al.}(2016)\citenamefont
  {Shainline}, \citenamefont {Buckley}, \citenamefont {Mirin},\ and\
  \citenamefont {{Sae Woo Nam}}}]{Shainline:16}%
  \BibitemOpen
  \bibfield  {author} {\bibinfo {author} {\bibfnamefont {J.~M.}\ \bibnamefont
  {Shainline}}, \bibinfo {author} {\bibfnamefont {S.~M.}\ \bibnamefont
  {Buckley}}, \bibinfo {author} {\bibfnamefont {R.~P.}\ \bibnamefont {Mirin}},
  \ and\ \bibinfo {author} {\bibnamefont {{Sae Woo Nam}}},\ }\href@noop {}
  {\bibfield  {journal} {\bibinfo  {journal} {arXiv preprint arXiv:1610.00053}\
  } (\bibinfo {year} {2016})}\BibitemShut {NoStop}%
\bibitem [{\citenamefont {Nahmias}\ \emph {et~al.}(2013)\citenamefont
  {Nahmias}, \citenamefont {Shastri}, \citenamefont {Tait},\ and\ \citenamefont
  {Prucnal}}]{nahmias2013leaky}%
  \BibitemOpen
  \bibfield  {author} {\bibinfo {author} {\bibfnamefont {M.~A.}\ \bibnamefont
  {Nahmias}}, \bibinfo {author} {\bibfnamefont {B.~J.}\ \bibnamefont
  {Shastri}}, \bibinfo {author} {\bibfnamefont {A.~N.}\ \bibnamefont {Tait}}, \
  and\ \bibinfo {author} {\bibfnamefont {P.~R.}\ \bibnamefont {Prucnal}},\
  }\href {\doibase 10.1109/JSTQE.2013.2257700} {\bibfield  {journal} {\bibinfo
  {journal} {IEEE J. Sel. Top. Quantum Electron.}\ }\textbf {\bibinfo {volume}
  {19}},\ \bibinfo {pages} {1} (\bibinfo {year} {2013})}\BibitemShut {NoStop}%
\bibitem [{\citenamefont {Prucnal}\ \emph {et~al.}(2016)\citenamefont
  {Prucnal}, \citenamefont {Shastri}, \citenamefont {{Ferreira de Lima}},
  \citenamefont {Nahmias},\ and\ \citenamefont {Tait}}]{Prucnal:16advances}%
  \BibitemOpen
  \bibfield  {author} {\bibinfo {author} {\bibfnamefont {P.~R.}\ \bibnamefont
  {Prucnal}}, \bibinfo {author} {\bibfnamefont {B.~J.}\ \bibnamefont
  {Shastri}}, \bibinfo {author} {\bibfnamefont {T.}~\bibnamefont {{Ferreira de
  Lima}}}, \bibinfo {author} {\bibfnamefont {M.~A.}\ \bibnamefont {Nahmias}}, \
  and\ \bibinfo {author} {\bibfnamefont {A.~N.}\ \bibnamefont {Tait}},\ }\href
  {\doibase 10.1364/AOP.8.000228} {\bibfield  {journal} {\bibinfo  {journal}
  {Adv. Opt. Photon.}\ }\textbf {\bibinfo {volume} {8}},\ \bibinfo {pages}
  {228} (\bibinfo {year} {2016})}\BibitemShut {NoStop}%
\bibitem [{\citenamefont {Selmi}\ \emph {et~al.}(2014)\citenamefont {Selmi},
  \citenamefont {Braive}, \citenamefont {Beaudoin}, \citenamefont {Sagnes},
  \citenamefont {Kuszelewicz},\ and\ \citenamefont {Barbay}}]{Selmi:2014}%
  \BibitemOpen
  \bibfield  {author} {\bibinfo {author} {\bibfnamefont {F.}~\bibnamefont
  {Selmi}}, \bibinfo {author} {\bibfnamefont {R.}~\bibnamefont {Braive}},
  \bibinfo {author} {\bibfnamefont {G.}~\bibnamefont {Beaudoin}}, \bibinfo
  {author} {\bibfnamefont {I.}~\bibnamefont {Sagnes}}, \bibinfo {author}
  {\bibfnamefont {R.}~\bibnamefont {Kuszelewicz}}, \ and\ \bibinfo {author}
  {\bibfnamefont {S.}~\bibnamefont {Barbay}},\ }\href {\doibase
  10.1103/PhysRevLett.112.183902} {\bibfield  {journal} {\bibinfo  {journal}
  {Phys. Rev. Lett.}\ }\textbf {\bibinfo {volume} {112}},\ \bibinfo {pages}
  {183902} (\bibinfo {year} {2014})}\BibitemShut {NoStop}%
\bibitem [{\citenamefont {Romeira}\ \emph {et~al.}(2016)\citenamefont
  {Romeira}, \citenamefont {Av\'{o}}, \citenamefont {Figueiredo}, \citenamefont
  {Barland},\ and\ \citenamefont {Javaloyes}}]{Romeira:2016}%
  \BibitemOpen
  \bibfield  {author} {\bibinfo {author} {\bibfnamefont {B.}~\bibnamefont
  {Romeira}}, \bibinfo {author} {\bibfnamefont {R.}~\bibnamefont {Av\'{o}}},
  \bibinfo {author} {\bibfnamefont {J.~M.~L.}\ \bibnamefont {Figueiredo}},
  \bibinfo {author} {\bibfnamefont {S.}~\bibnamefont {Barland}}, \ and\
  \bibinfo {author} {\bibfnamefont {J.}~\bibnamefont {Javaloyes}},\ }\href
  {\doibase 10.1038/srep19510} {\bibfield  {journal} {\bibinfo  {journal}
  {Scientific Reports}\ }\textbf {\bibinfo {volume} {6}},\ \bibinfo {pages}
  {19510 EP } (\bibinfo {year} {2016})}\BibitemShut {NoStop}%
\bibitem [{\citenamefont {Nahmias}\ \emph {et~al.}(2016)\citenamefont
  {Nahmias}, \citenamefont {Tait}, \citenamefont {Tolias}, \citenamefont
  {Chang}, \citenamefont {{Ferreira de Lima}}, \citenamefont {Shastri},\ and\
  \citenamefont {Prucnal}}]{Nahmias:16}%
  \BibitemOpen
  \bibfield  {author} {\bibinfo {author} {\bibfnamefont {M.~A.}\ \bibnamefont
  {Nahmias}}, \bibinfo {author} {\bibfnamefont {A.~N.}\ \bibnamefont {Tait}},
  \bibinfo {author} {\bibfnamefont {L.}~\bibnamefont {Tolias}}, \bibinfo
  {author} {\bibfnamefont {M.~P.}\ \bibnamefont {Chang}}, \bibinfo {author}
  {\bibfnamefont {T.}~\bibnamefont {{Ferreira de Lima}}}, \bibinfo {author}
  {\bibfnamefont {B.~J.}\ \bibnamefont {Shastri}}, \ and\ \bibinfo {author}
  {\bibfnamefont {P.~R.}\ \bibnamefont {Prucnal}},\ }\href {\doibase
  http://dx.doi.org/10.1063/1.4945368} {\bibfield  {journal} {\bibinfo
  {journal} {Applied Physics Letters}\ }\textbf {\bibinfo {volume} {108}},\
  \bibinfo {eid} {151106} (\bibinfo {year} {2016}),\
  http://dx.doi.org/10.1063/1.4945368}\BibitemShut {NoStop}%
\bibitem [{\citenamefont {Vaerenbergh}\ \emph {et~al.}(2012)\citenamefont
  {Vaerenbergh}, \citenamefont {Fiers}, \citenamefont {Mechet}, \citenamefont
  {Spuesens}, \citenamefont {Kumar}, \citenamefont {Morthier}, \citenamefont
  {Schrauwen}, \citenamefont {Dambre},\ and\ \citenamefont
  {Bienstman}}]{VanVaerenbergh:12}%
  \BibitemOpen
  \bibfield  {author} {\bibinfo {author} {\bibfnamefont {T.~V.}\ \bibnamefont
  {Vaerenbergh}}, \bibinfo {author} {\bibfnamefont {M.}~\bibnamefont {Fiers}},
  \bibinfo {author} {\bibfnamefont {P.}~\bibnamefont {Mechet}}, \bibinfo
  {author} {\bibfnamefont {T.}~\bibnamefont {Spuesens}}, \bibinfo {author}
  {\bibfnamefont {R.}~\bibnamefont {Kumar}}, \bibinfo {author} {\bibfnamefont
  {G.}~\bibnamefont {Morthier}}, \bibinfo {author} {\bibfnamefont
  {B.}~\bibnamefont {Schrauwen}}, \bibinfo {author} {\bibfnamefont
  {J.}~\bibnamefont {Dambre}}, \ and\ \bibinfo {author} {\bibfnamefont
  {P.}~\bibnamefont {Bienstman}},\ }\href {\doibase 10.1364/OE.20.020292}
  {\bibfield  {journal} {\bibinfo  {journal} {Opt. Express}\ }\textbf {\bibinfo
  {volume} {20}},\ \bibinfo {pages} {20292} (\bibinfo {year}
  {2012})}\BibitemShut {NoStop}%
\bibitem [{\citenamefont {Shastri}\ \emph {et~al.}(2015)\citenamefont
  {Shastri}, \citenamefont {Nahmias}, \citenamefont {Tait}, \citenamefont
  {Rodriguez}, \citenamefont {Wu},\ and\ \citenamefont
  {Prucnal}}]{Shastri:2015}%
  \BibitemOpen
  \bibfield  {author} {\bibinfo {author} {\bibfnamefont {B.~J.}\ \bibnamefont
  {Shastri}}, \bibinfo {author} {\bibfnamefont {M.~A.}\ \bibnamefont
  {Nahmias}}, \bibinfo {author} {\bibfnamefont {A.~N.}\ \bibnamefont {Tait}},
  \bibinfo {author} {\bibfnamefont {A.~W.}\ \bibnamefont {Rodriguez}}, \bibinfo
  {author} {\bibfnamefont {B.}~\bibnamefont {Wu}}, \ and\ \bibinfo {author}
  {\bibfnamefont {P.~R.}\ \bibnamefont {Prucnal}},\ }\href {\doibase
  10.1038/srep19126} {\bibfield  {journal} {\bibinfo  {journal} {Sci. Rep.}\
  }\textbf {\bibinfo {volume} {5}},\ \bibinfo {pages} {19126} (\bibinfo {year}
  {2015})}\BibitemShut {NoStop}%
\bibitem [{\citenamefont {Zhang}\ \emph {et~al.}(2012)\citenamefont {Zhang},
  \citenamefont {Virally}, \citenamefont {Bao}, \citenamefont {Ping},
  \citenamefont {Massar}, \citenamefont {Godbout},\ and\ \citenamefont
  {Kockaert}}]{Zhang:12}%
  \BibitemOpen
  \bibfield  {author} {\bibinfo {author} {\bibfnamefont {H.}~\bibnamefont
  {Zhang}}, \bibinfo {author} {\bibfnamefont {S.}~\bibnamefont {Virally}},
  \bibinfo {author} {\bibfnamefont {Q.}~\bibnamefont {Bao}}, \bibinfo {author}
  {\bibfnamefont {L.~K.}\ \bibnamefont {Ping}}, \bibinfo {author}
  {\bibfnamefont {S.}~\bibnamefont {Massar}}, \bibinfo {author} {\bibfnamefont
  {N.}~\bibnamefont {Godbout}}, \ and\ \bibinfo {author} {\bibfnamefont
  {P.}~\bibnamefont {Kockaert}},\ }\href {\doibase 10.1364/OL.37.001856}
  {\bibfield  {journal} {\bibinfo  {journal} {Opt. Lett.}\ }\textbf {\bibinfo
  {volume} {37}},\ \bibinfo {pages} {1856} (\bibinfo {year}
  {2012})}\BibitemShut {NoStop}%
\bibitem [{\citenamefont {Hill}\ \emph {et~al.}(2002)\citenamefont {Hill},
  \citenamefont {Frietman}, \citenamefont {de~Waardt}, \citenamefont {Khoe},\
  and\ \citenamefont {Dorren}}]{Hill:2002}%
  \BibitemOpen
  \bibfield  {author} {\bibinfo {author} {\bibfnamefont {M.}~\bibnamefont
  {Hill}}, \bibinfo {author} {\bibfnamefont {E.~E.~E.}\ \bibnamefont
  {Frietman}}, \bibinfo {author} {\bibfnamefont {H.}~\bibnamefont {de~Waardt}},
  \bibinfo {author} {\bibfnamefont {G.-D.}\ \bibnamefont {Khoe}}, \ and\
  \bibinfo {author} {\bibfnamefont {H.}~\bibnamefont {Dorren}},\ }\href
  {\doibase 10.1109/TNN.2002.804222} {\bibfield  {journal} {\bibinfo  {journal}
  {IEEE Trans. Neural Networks}\ }\textbf {\bibinfo {volume} {13}},\ \bibinfo
  {pages} {1504} (\bibinfo {year} {2002})}\BibitemShut {NoStop}%
\bibitem [{\citenamefont {Brunner}\ \emph {et~al.}(2013)\citenamefont
  {Brunner}, \citenamefont {Soriano}, \citenamefont {Mirasso},\ and\
  \citenamefont {Fischer}}]{Brunner:2013}%
  \BibitemOpen
  \bibfield  {author} {\bibinfo {author} {\bibfnamefont {D.}~\bibnamefont
  {Brunner}}, \bibinfo {author} {\bibfnamefont {M.~C.}\ \bibnamefont
  {Soriano}}, \bibinfo {author} {\bibfnamefont {C.~R.}\ \bibnamefont
  {Mirasso}}, \ and\ \bibinfo {author} {\bibfnamefont {I.}~\bibnamefont
  {Fischer}},\ }\href {\doibase 10.1038/ncomms2368} {\bibfield  {journal}
  {\bibinfo  {journal} {Nat Commun}\ }\textbf {\bibinfo {volume} {4}},\
  \bibinfo {pages} {1364} (\bibinfo {year} {2013})}\BibitemShut {NoStop}%
\bibitem [{\citenamefont {Vandoorne}\ \emph {et~al.}(2014)\citenamefont
  {Vandoorne}, \citenamefont {Mechet}, \citenamefont {Van~Vaerenbergh},
  \citenamefont {Fiers}, \citenamefont {Morthier}, \citenamefont {Verstraeten},
  \citenamefont {Schrauwen}, \citenamefont {Dambre},\ and\ \citenamefont
  {Bienstman}}]{Vandoorne:2014}%
  \BibitemOpen
  \bibfield  {author} {\bibinfo {author} {\bibfnamefont {K.}~\bibnamefont
  {Vandoorne}}, \bibinfo {author} {\bibfnamefont {P.}~\bibnamefont {Mechet}},
  \bibinfo {author} {\bibfnamefont {T.}~\bibnamefont {Van~Vaerenbergh}},
  \bibinfo {author} {\bibfnamefont {M.}~\bibnamefont {Fiers}}, \bibinfo
  {author} {\bibfnamefont {G.}~\bibnamefont {Morthier}}, \bibinfo {author}
  {\bibfnamefont {D.}~\bibnamefont {Verstraeten}}, \bibinfo {author}
  {\bibfnamefont {B.}~\bibnamefont {Schrauwen}}, \bibinfo {author}
  {\bibfnamefont {J.}~\bibnamefont {Dambre}}, \ and\ \bibinfo {author}
  {\bibfnamefont {P.}~\bibnamefont {Bienstman}},\ }\href {\doibase
  10.1038/ncomms4541} {\bibfield  {journal} {\bibinfo  {journal} {Nat Commun}\
  }\textbf {\bibinfo {volume} {5}} (\bibinfo {year} {2014}),\
  10.1038/ncomms4541}\BibitemShut {NoStop}%
\bibitem [{\citenamefont {Soriano}\ \emph {et~al.}(2015)\citenamefont
  {Soriano}, \citenamefont {Brunner}, \citenamefont {Escalona-Mor{\'a}n},
  \citenamefont {Mirasso},\ and\ \citenamefont {Fischer}}]{Soriano:15}%
  \BibitemOpen
  \bibfield  {author} {\bibinfo {author} {\bibfnamefont {M.~C.}\ \bibnamefont
  {Soriano}}, \bibinfo {author} {\bibfnamefont {D.}~\bibnamefont {Brunner}},
  \bibinfo {author} {\bibfnamefont {M.}~\bibnamefont {Escalona-Mor{\'a}n}},
  \bibinfo {author} {\bibfnamefont {C.~R.}\ \bibnamefont {Mirasso}}, \ and\
  \bibinfo {author} {\bibfnamefont {I.}~\bibnamefont {Fischer}},\ }\href
  {\doibase 10.3389/fncom.2015.00068} {\bibfield  {journal} {\bibinfo
  {journal} {Frontiers in Computational Neuroscience}\ }\textbf {\bibinfo
  {volume} {9}},\ \bibinfo {pages} {68} (\bibinfo {year} {2015})}\BibitemShut
  {NoStop}%
\bibitem [{\citenamefont {Duport}\ \emph {et~al.}(2016)\citenamefont {Duport},
  \citenamefont {Smerieri}, \citenamefont {Akrout}, \citenamefont
  {Haelterman},\ and\ \citenamefont {Massar}}]{Duport:2016}%
  \BibitemOpen
  \bibfield  {author} {\bibinfo {author} {\bibfnamefont {F.}~\bibnamefont
  {Duport}}, \bibinfo {author} {\bibfnamefont {A.}~\bibnamefont {Smerieri}},
  \bibinfo {author} {\bibfnamefont {A.}~\bibnamefont {Akrout}}, \bibinfo
  {author} {\bibfnamefont {M.}~\bibnamefont {Haelterman}}, \ and\ \bibinfo
  {author} {\bibfnamefont {S.}~\bibnamefont {Massar}},\ }\href {\doibase
  10.1038/srep22381} {\bibfield  {journal} {\bibinfo  {journal} {Scientific
  Reports}\ }\textbf {\bibinfo {volume} {6}},\ \bibinfo {pages} {22381 EP }
  (\bibinfo {year} {2016})}\BibitemShut {NoStop}%
\bibitem [{\citenamefont {Vandoorne}\ \emph {et~al.}(2008)\citenamefont
  {Vandoorne}, \citenamefont {Dierckx}, \citenamefont {Schrauwen},
  \citenamefont {Verstraeten}, \citenamefont {Baets}, \citenamefont
  {Bienstman},\ and\ \citenamefont {Campenhout}}]{Vandoorne:08}%
  \BibitemOpen
  \bibfield  {author} {\bibinfo {author} {\bibfnamefont {K.}~\bibnamefont
  {Vandoorne}}, \bibinfo {author} {\bibfnamefont {W.}~\bibnamefont {Dierckx}},
  \bibinfo {author} {\bibfnamefont {B.}~\bibnamefont {Schrauwen}}, \bibinfo
  {author} {\bibfnamefont {D.}~\bibnamefont {Verstraeten}}, \bibinfo {author}
  {\bibfnamefont {R.}~\bibnamefont {Baets}}, \bibinfo {author} {\bibfnamefont
  {P.}~\bibnamefont {Bienstman}}, \ and\ \bibinfo {author} {\bibfnamefont
  {J.~V.}\ \bibnamefont {Campenhout}},\ }\href {\doibase 10.1364/OE.16.011182}
  {\bibfield  {journal} {\bibinfo  {journal} {Opt. Express}\ }\textbf {\bibinfo
  {volume} {16}},\ \bibinfo {pages} {11182} (\bibinfo {year}
  {2008})}\BibitemShut {NoStop}%
\bibitem [{\citenamefont {Mesaritakis}\ \emph {et~al.}(2013)\citenamefont
  {Mesaritakis}, \citenamefont {Papataxiarhis},\ and\ \citenamefont
  {Syvridis}}]{Mesaritakis:13}%
  \BibitemOpen
  \bibfield  {author} {\bibinfo {author} {\bibfnamefont {C.}~\bibnamefont
  {Mesaritakis}}, \bibinfo {author} {\bibfnamefont {V.}~\bibnamefont
  {Papataxiarhis}}, \ and\ \bibinfo {author} {\bibfnamefont {D.}~\bibnamefont
  {Syvridis}},\ }\href {\doibase 10.1364/JOSAB.30.003048} {\bibfield  {journal}
  {\bibinfo  {journal} {J. Opt. Soc. Am. B}\ }\textbf {\bibinfo {volume}
  {30}},\ \bibinfo {pages} {3048} (\bibinfo {year} {2013})}\BibitemShut
  {NoStop}%
\bibitem [{\citenamefont {Hopfield}\ and\ \citenamefont
  {Tank}(1985)}]{Hopfield:85}%
  \BibitemOpen
  \bibfield  {author} {\bibinfo {author} {\bibfnamefont {J.~J.}\ \bibnamefont
  {Hopfield}}\ and\ \bibinfo {author} {\bibfnamefont {D.~W.}\ \bibnamefont
  {Tank}},\ }\href {\doibase 10.1007/BF00339943} {\bibfield  {journal}
  {\bibinfo  {journal} {Biological Cybernetics}\ }\textbf {\bibinfo {volume}
  {52}},\ \bibinfo {pages} {141} (\bibinfo {year} {1985})}\BibitemShut
  {NoStop}%
\bibitem [{\citenamefont {Tumuluru}\ \emph {et~al.}(2010)\citenamefont
  {Tumuluru}, \citenamefont {Wang},\ and\ \citenamefont
  {Niyato}}]{Tumuluru:10}%
  \BibitemOpen
  \bibfield  {author} {\bibinfo {author} {\bibfnamefont {V.~K.}\ \bibnamefont
  {Tumuluru}}, \bibinfo {author} {\bibfnamefont {P.}~\bibnamefont {Wang}}, \
  and\ \bibinfo {author} {\bibfnamefont {D.}~\bibnamefont {Niyato}},\ }in\
  \href {\doibase 10.1109/ICC.2010.5502348} {\emph {\bibinfo {booktitle}
  {Communications (ICC), 2010 IEEE International Conference on}}}\ (\bibinfo
  {year} {2010})\ pp.\ \bibinfo {pages} {1--5}\BibitemShut {NoStop}%
\bibitem [{\citenamefont {Mitra}\ and\ \citenamefont {Poor}(1994)}]{Mitra:94}%
  \BibitemOpen
  \bibfield  {author} {\bibinfo {author} {\bibfnamefont {U.}~\bibnamefont
  {Mitra}}\ and\ \bibinfo {author} {\bibfnamefont {H.~V.}\ \bibnamefont
  {Poor}},\ }\href {\doibase 10.1109/49.339913} {\bibfield  {journal} {\bibinfo
   {journal} {IEEE Journal on Selected Areas in Communications}\ }\textbf
  {\bibinfo {volume} {12}},\ \bibinfo {pages} {1460} (\bibinfo {year}
  {1994})}\BibitemShut {NoStop}%
\bibitem [{\citenamefont {Du}\ \emph {et~al.}(2002)\citenamefont {Du},
  \citenamefont {Lai}, \citenamefont {Cheng},\ and\ \citenamefont
  {Swamy}}]{Du:02}%
  \BibitemOpen
  \bibfield  {author} {\bibinfo {author} {\bibfnamefont {K.-L.}\ \bibnamefont
  {Du}}, \bibinfo {author} {\bibfnamefont {A.}~\bibnamefont {Lai}}, \bibinfo
  {author} {\bibfnamefont {K.}~\bibnamefont {Cheng}}, \ and\ \bibinfo {author}
  {\bibfnamefont {M.}~\bibnamefont {Swamy}},\ }\href {\doibase
  http://dx.doi.org/10.1016/S0165-1684(01)00185-2} {\bibfield  {journal}
  {\bibinfo  {journal} {Signal Processing}\ }\textbf {\bibinfo {volume} {82}},\
  \bibinfo {pages} {547 } (\bibinfo {year} {2002})}\BibitemShut {NoStop}%
\bibitem [{\citenamefont {Tait}\ \emph {et~al.}(2017)\citenamefont {Tait},
  \citenamefont {{Ferreira de Lima}}, \citenamefont {Wu}, \citenamefont {Zhou},
  \citenamefont {Chang}, \citenamefont {Nahmias}, \citenamefont {Shastri},\
  and\ \citenamefont {Prucnal}}]{Tait:17cleo}%
  \BibitemOpen
  \bibfield  {author} {\bibinfo {author} {\bibfnamefont {A.}~\bibnamefont
  {Tait}}, \bibinfo {author} {\bibfnamefont {T.}~\bibnamefont {{Ferreira de
  Lima}}}, \bibinfo {author} {\bibfnamefont {A.}~\bibnamefont {Wu}}, \bibinfo
  {author} {\bibfnamefont {E.}~\bibnamefont {Zhou}}, \bibinfo {author}
  {\bibfnamefont {M.}~\bibnamefont {Chang}}, \bibinfo {author} {\bibfnamefont
  {M.}~\bibnamefont {Nahmias}}, \bibinfo {author} {\bibfnamefont
  {B.}~\bibnamefont {Shastri}}, \ and\ \bibinfo {author} {\bibfnamefont
  {P.}~\bibnamefont {Prucnal}},\ }in\ \href@noop {} {\emph {\bibinfo
  {booktitle} {CLEO: 2017}}}\ (\bibinfo  {publisher} {IEEE},\ \bibinfo {year}
  {2017})\BibitemShut {NoStop}%
\bibitem [{\citenamefont {Bojko}\ \emph {et~al.}(2011)\citenamefont {Bojko},
  \citenamefont {Li}, \citenamefont {He}, \citenamefont {Baehr-Jones},
  \citenamefont {Hochberg},\ and\ \citenamefont {Aida}}]{Bojko:11}%
  \BibitemOpen
  \bibfield  {author} {\bibinfo {author} {\bibfnamefont {R.~J.}\ \bibnamefont
  {Bojko}}, \bibinfo {author} {\bibfnamefont {J.}~\bibnamefont {Li}}, \bibinfo
  {author} {\bibfnamefont {L.}~\bibnamefont {He}}, \bibinfo {author}
  {\bibfnamefont {T.}~\bibnamefont {Baehr-Jones}}, \bibinfo {author}
  {\bibfnamefont {M.}~\bibnamefont {Hochberg}}, \ and\ \bibinfo {author}
  {\bibfnamefont {Y.}~\bibnamefont {Aida}},\ }\href {\doibase
  http://dx.doi.org/10.1116/1.3653266} {\bibfield  {journal} {\bibinfo
  {journal} {J. Vac. Sci. Technol., B}\ }\textbf {\bibinfo {volume} {29}}
  (\bibinfo {year} {2011}),\ http://dx.doi.org/10.1116/1.3653266}\BibitemShut
  {NoStop}%
\bibitem [{\citenamefont {Wang}\ \emph {et~al.}(2014)\citenamefont {Wang},
  \citenamefont {Wang}, \citenamefont {Flueckiger}, \citenamefont {Yun},
  \citenamefont {Shi}, \citenamefont {Bojko}, \citenamefont {Jaeger},\ and\
  \citenamefont {Chrostowski}}]{Wang:14opex}%
  \BibitemOpen
  \bibfield  {author} {\bibinfo {author} {\bibfnamefont {Y.}~\bibnamefont
  {Wang}}, \bibinfo {author} {\bibfnamefont {X.}~\bibnamefont {Wang}}, \bibinfo
  {author} {\bibfnamefont {J.}~\bibnamefont {Flueckiger}}, \bibinfo {author}
  {\bibfnamefont {H.}~\bibnamefont {Yun}}, \bibinfo {author} {\bibfnamefont
  {W.}~\bibnamefont {Shi}}, \bibinfo {author} {\bibfnamefont {R.}~\bibnamefont
  {Bojko}}, \bibinfo {author} {\bibfnamefont {N.~A.}\ \bibnamefont {Jaeger}}, \
  and\ \bibinfo {author} {\bibfnamefont {L.}~\bibnamefont {Chrostowski}},\
  }\href {\doibase 10.1364/OE.22.020652} {\bibfield  {journal} {\bibinfo
  {journal} {Opt. Express}\ }\textbf {\bibinfo {volume} {22}},\ \bibinfo
  {pages} {20652} (\bibinfo {year} {2014})}\BibitemShut {NoStop}%
\bibitem [{\citenamefont {Tait}\ \emph
  {et~al.}(2016{\natexlab{c}})\citenamefont {Tait}, \citenamefont {{Ferreira de
  Lima}}, \citenamefont {Nahmias}, \citenamefont {Shastri},\ and\ \citenamefont
  {Prucnal}}]{Tait:15cont}%
  \BibitemOpen
  \bibfield  {author} {\bibinfo {author} {\bibfnamefont {A.}~\bibnamefont
  {Tait}}, \bibinfo {author} {\bibfnamefont {T.}~\bibnamefont {{Ferreira de
  Lima}}}, \bibinfo {author} {\bibfnamefont {M.}~\bibnamefont {Nahmias}},
  \bibinfo {author} {\bibfnamefont {B.}~\bibnamefont {Shastri}}, \ and\
  \bibinfo {author} {\bibfnamefont {P.}~\bibnamefont {Prucnal}},\ }\href
  {\doibase 10.1109/LPT.2016.2516440} {\bibfield  {journal} {\bibinfo
  {journal} {Photonics Technol. Lett.}\ }\textbf {\bibinfo {volume} {28}},\
  \bibinfo {pages} {887} (\bibinfo {year} {2016}{\natexlab{c}})}\BibitemShut
  {NoStop}%
\bibitem [{\citenamefont {Zhang}\ \emph {et~al.}(2013)\citenamefont {Zhang},
  \citenamefont {Yang}, \citenamefont {Lim}, \citenamefont {Lo}, \citenamefont
  {Galland}, \citenamefont {Baehr-Jones},\ and\ \citenamefont
  {Hochberg}}]{Zhang:13}%
  \BibitemOpen
  \bibfield  {author} {\bibinfo {author} {\bibfnamefont {Y.}~\bibnamefont
  {Zhang}}, \bibinfo {author} {\bibfnamefont {S.}~\bibnamefont {Yang}},
  \bibinfo {author} {\bibfnamefont {A.~E.-J.}\ \bibnamefont {Lim}}, \bibinfo
  {author} {\bibfnamefont {G.-Q.}\ \bibnamefont {Lo}}, \bibinfo {author}
  {\bibfnamefont {C.}~\bibnamefont {Galland}}, \bibinfo {author} {\bibfnamefont
  {T.}~\bibnamefont {Baehr-Jones}}, \ and\ \bibinfo {author} {\bibfnamefont
  {M.}~\bibnamefont {Hochberg}},\ }\href {\doibase 10.1364/OE.21.001310}
  {\bibfield  {journal} {\bibinfo  {journal} {Opt. Express}\ }\textbf {\bibinfo
  {volume} {21}},\ \bibinfo {pages} {1310} (\bibinfo {year}
  {2013})}\BibitemShut {NoStop}%
\bibitem [{\citenamefont {Bekolay}\ \emph {et~al.}(2013)\citenamefont
  {Bekolay}, \citenamefont {Bergstra}, \citenamefont {Hunsberger},
  \citenamefont {DeWolf}, \citenamefont {Stewart}, \citenamefont {Rasmussen},
  \citenamefont {Choo}, \citenamefont {Voelker},\ and\ \citenamefont
  {Eliasmith}}]{Bekolay:2013}%
  \BibitemOpen
  \bibfield  {author} {\bibinfo {author} {\bibfnamefont {T.}~\bibnamefont
  {Bekolay}}, \bibinfo {author} {\bibfnamefont {J.}~\bibnamefont {Bergstra}},
  \bibinfo {author} {\bibfnamefont {E.}~\bibnamefont {Hunsberger}}, \bibinfo
  {author} {\bibfnamefont {T.}~\bibnamefont {DeWolf}}, \bibinfo {author}
  {\bibfnamefont {T.~C.}\ \bibnamefont {Stewart}}, \bibinfo {author}
  {\bibfnamefont {D.}~\bibnamefont {Rasmussen}}, \bibinfo {author}
  {\bibfnamefont {X.}~\bibnamefont {Choo}}, \bibinfo {author} {\bibfnamefont
  {A.~R.}\ \bibnamefont {Voelker}}, \ and\ \bibinfo {author} {\bibfnamefont
  {C.}~\bibnamefont {Eliasmith}},\ }\href {\doibase 10.3389/fninf.2013.00048}
  {\bibfield  {journal} {\bibinfo  {journal} {Frontiers in Neuroinformatics}\
  }\textbf {\bibinfo {volume} {7}},\ \bibinfo {pages} {48} (\bibinfo {year}
  {2013})}\BibitemShut {NoStop}%
\bibitem [{\citenamefont {Friedl}\ \emph {et~al.}(2016)\citenamefont {Friedl},
  \citenamefont {Voelker}, \citenamefont {Peer},\ and\ \citenamefont
  {Eliasmith}}]{Friedl:16}%
  \BibitemOpen
  \bibfield  {author} {\bibinfo {author} {\bibfnamefont {K.~E.}\ \bibnamefont
  {Friedl}}, \bibinfo {author} {\bibfnamefont {A.~R.}\ \bibnamefont {Voelker}},
  \bibinfo {author} {\bibfnamefont {A.}~\bibnamefont {Peer}}, \ and\ \bibinfo
  {author} {\bibfnamefont {C.}~\bibnamefont {Eliasmith}},\ }\href {\doibase
  10.1109/LRA.2016.2517213} {\bibfield  {journal} {\bibinfo  {journal} {IEEE
  Robotics and Automation Letters}\ }\textbf {\bibinfo {volume} {1}},\ \bibinfo
  {pages} {516} (\bibinfo {year} {2016})}\BibitemShut {NoStop}%
\bibitem [{\citenamefont {Mundy}\ \emph {et~al.}(2015)\citenamefont {Mundy},
  \citenamefont {Knight}, \citenamefont {Stewart},\ and\ \citenamefont
  {Furber}}]{Mundy:15}%
  \BibitemOpen
  \bibfield  {author} {\bibinfo {author} {\bibfnamefont {A.}~\bibnamefont
  {Mundy}}, \bibinfo {author} {\bibfnamefont {J.}~\bibnamefont {Knight}},
  \bibinfo {author} {\bibfnamefont {T.}~\bibnamefont {Stewart}}, \ and\
  \bibinfo {author} {\bibfnamefont {S.}~\bibnamefont {Furber}},\ }in\ \href
  {\doibase 10.1109/IJCNN.2015.7280390} {\emph {\bibinfo {booktitle} {Neural
  Networks (IJCNN), 2015 International Joint Conference on}}}\ (\bibinfo {year}
  {2015})\ pp.\ \bibinfo {pages} {1--8}\BibitemShut {NoStop}%
\bibitem [{\citenamefont {Chrostowski}\ \emph {et~al.}(2014)\citenamefont
  {Chrostowski}, \citenamefont {Wang}, \citenamefont {Flueckiger},
  \citenamefont {Wu}, \citenamefont {Wang},\ and\ \citenamefont
  {Fard}}]{Chrostowski:14}%
  \BibitemOpen
  \bibfield  {author} {\bibinfo {author} {\bibfnamefont {L.}~\bibnamefont
  {Chrostowski}}, \bibinfo {author} {\bibfnamefont {X.}~\bibnamefont {Wang}},
  \bibinfo {author} {\bibfnamefont {J.}~\bibnamefont {Flueckiger}}, \bibinfo
  {author} {\bibfnamefont {Y.}~\bibnamefont {Wu}}, \bibinfo {author}
  {\bibfnamefont {Y.}~\bibnamefont {Wang}}, \ and\ \bibinfo {author}
  {\bibfnamefont {S.~T.}\ \bibnamefont {Fard}},\ }in\ \href {\doibase
  10.1364/OFC.2014.Th2A.37} {\emph {\bibinfo {booktitle} {Optical Fiber
  Communication Conference}}}\ (\bibinfo  {publisher} {Optical Society of
  America},\ \bibinfo {year} {2014})\ p.\ \bibinfo {pages}
  {Th2A.37}\BibitemShut {NoStop}%
\bibitem [{\citenamefont {Jayatilleka}\ \emph {et~al.}(2015)\citenamefont
  {Jayatilleka}, \citenamefont {Murray}, \citenamefont {\'{A}ngel
  Guill\'{e}n-Torres}, \citenamefont {Caverley}, \citenamefont {Hu},
  \citenamefont {Jaeger}, \citenamefont {Chrostowski},\ and\ \citenamefont
  {Shekhar}}]{Jayatilleka:15opex}%
  \BibitemOpen
  \bibfield  {author} {\bibinfo {author} {\bibfnamefont {H.}~\bibnamefont
  {Jayatilleka}}, \bibinfo {author} {\bibfnamefont {K.}~\bibnamefont {Murray}},
  \bibinfo {author} {\bibfnamefont {M.}~\bibnamefont {\'{A}ngel
  Guill\'{e}n-Torres}}, \bibinfo {author} {\bibfnamefont {M.}~\bibnamefont
  {Caverley}}, \bibinfo {author} {\bibfnamefont {R.}~\bibnamefont {Hu}},
  \bibinfo {author} {\bibfnamefont {N.~A.~F.}\ \bibnamefont {Jaeger}}, \bibinfo
  {author} {\bibfnamefont {L.}~\bibnamefont {Chrostowski}}, \ and\ \bibinfo
  {author} {\bibfnamefont {S.}~\bibnamefont {Shekhar}},\ }\href {\doibase
  10.1364/OE.23.025084} {\bibfield  {journal} {\bibinfo  {journal} {Opt.
  Express}\ }\textbf {\bibinfo {volume} {23}},\ \bibinfo {pages} {25084}
  (\bibinfo {year} {2015})}\BibitemShut {NoStop}%
\end{thebibliography}

%

\end{document}